\documentclass[10pt, final]{IEEEtran}

\usepackage{amsthm}

\usepackage{amssymb}
\usepackage{amsmath}
\usepackage{bbm}
\usepackage{mathrsfs}
\usepackage{cite}
\usepackage{bm,epsfig,amsthm,url}
\usepackage{indentfirst}
\usepackage{amsfonts}
\usepackage{epstopdf}
\usepackage{cases}
\usepackage[caption=false,font=scriptsize]{subfig}
\usepackage{xcolor}
\usepackage{mathtools}

\makeatletter
\renewcommand{\maketag@@@}[1]{\hbox{\m@th\normalsize\normalfont#1}}%
\makeatother
\usepackage{hyperref}
\usepackage{stfloats}

\newtheoremstyle{mystyle}{}{}{}{}{}{: }{0pt}{\indent \it{\thmname{#1}\thmnumber{ #2}\thmnote{#3}}}
\theoremstyle{mystyle}

\newtheorem{Proposition}{Proposition}

\newtheorem{Remark}{Remark}

\usepackage[noend]{algpseudocode}
\usepackage{algorithmicx,algorithm}
\usepackage{float}

\begin{document}

\title{Semi-Passive Intelligent Reflecting Surface Enabled Sensing Systems}

\author{{Qiaoyan~Peng, Qingqing~Wu,~\IEEEmembership{Senior Member, IEEE},~Wen~Chen,~\IEEEmembership{Senior Member, IEEE},~Shaodan~Ma,~\IEEEmembership{Senior Member, IEEE},~Ming-Min Zhao,~\IEEEmembership{Member, IEEE},~and~Octavia A. Dobre,~\IEEEmembership{Fellow, IEEE}}
\thanks{Qiaoyan Peng is with the State Key Laboratory of Internet of Things for Smart City, University of Macau, Macao 999078, China, and also with the Institute for Signal Processing and Systems at Shanghai Jiao Tong University, 200240, China (email: yc27464@um.edu.mo).

Qingqing~Wu and Wen~Chen are with the
Department of Electronic Engineering, Shanghai Jiao Tong University, Shanghai 200240, China (e-mail: qingqingwu@sjtu.edu.cn; wenchen@sjtu.edu.cn).

Shaodan Ma is with the State Key Laboratory of Internet of Things for Smart City, University of Macau, Macao 999078, China (email: shaodanma@um.edu.mo).

Ming-Min Zhao is with the College of Information Science and Electronic Engineering, Zhejiang University, China (e-mail: zmmblack@zju.edu.cn).

Octavia A. Dobre is with the Faculty of Engineering and Applied Science, Memorial University, St. John’s, NL A1B 3X5, Canada (e-mail: odobre@mun.ca).
}
}

\maketitle

\begin{abstract}
Intelligent reflecting surface (IRS) has garnered growing interest and attention due to its potential for facilitating and supporting wireless communications and sensing. This paper studies a semi-passive IRS-enabled sensing system, where an IRS consists of both passive reflecting elements and active sensors. Our goal is to minimize the Cram\'{e}r-Rao bound (CRB) for parameter estimation under both point and extended target cases. Towards this goal, we begin by deriving the CRB for the direction-of-arrival (DoA) estimation in closed-form and then theoretically analyze the IRS reflecting elements and sensors allocation design based on the CRB under the point target case with a single-antenna base station (BS). To efficiently solve the corresponding optimization problem for the case with a multi-antenna BS, we propose an efficient algorithm by jointly optimizing the IRS phase shifts and the BS beamformers. Under the extended target case, the CRB for the target response matrix (TRM) estimation is minimized via the optimization of the BS transmit beamformers. Moreover, we explore the influence of various system parameters on the CRB and compare these effects to those observed under the point target case. Simulation results show the effectiveness of the semi-passive IRS and our proposed beamforming design for improving the performance of the sensing system.
\end{abstract}
\begin{IEEEkeywords}
Intelligent reflecting surfaces, semi-passive IRS architecture, Cram\'{e}r-Rao bound, direction-of-arrival, target response matrix, beamforming, wireless sensing.
\end{IEEEkeywords}

\section{Introduction}
Due to the emergence of environment-aware applications, including autonomous driving, unmanned aerial vehicle tracking, and human activity recognition, sensing services need to fulfill increasingly stringent demands \cite{auto,UAV,human,6G,wu2023intelligent,Oct2020}. These applications require not only reliable communications but also advanced sensing capabilities to collect real-time information on the surrounding environment. In massive multi-input multi-output systems, a large number of antennas are deployed at the transmitter (TX) and receiver (RX), thereby improving spatial resolution and received power. In addition, millimeter-wave/terahertz technologies offer higher carrier frequencies, thereby providing wider bandwidth and higher spatial resolution. With the development of these technologies, a base station (BS) or an access point can achieve high-reliable and high-accuracy sensing \cite{MIMO,mmWave,THz}. Moreover, the existing communication infrastructures can be easily reutilized for sensing with adjustments to hardware, signaling strategies, and communication standards \cite{infrastructures}. Without dedicated sensing infrastructures, it provides new opportunities for cost-effective and simplified network architectures for sensing services.

Wireless/radio frequency sensing refers to the use of transmission, reflection, diffraction, and/or scattering of radio waves to detect the presence of objects in the environment or to estimate their various physical properties, e.g., position, direction, and velocity \cite{RFsensing}. In traditional wireless sensing/localization, line-of-sight (LoS) paths are exploited to extract the sensing information from the echo signals \cite{AP_rely_LoS}. Specifically, in a mono-static BS sensing system, the TXs and RXs are located at the same location, or the BS transmits and receives the probing signals for a full-duplex mode \cite{mono-static}. In contrast, sensing is conducted at the RX, which is situated at a different location from the TX in a bi-static BS sensing system \cite{bi-static}. For sensing, the non-LoS (NLoS) path is generally considered as interference that can negatively impact the performance of target sensing \cite{kaitao}. Therefore, the sensing performance of both mono-static and bi-static BS sensing systems is limited due to the lack of LoS links between the BS and targets caused by dense obstacles as well as significant signal attenuation across long distances.
 
The intelligent reflecting surface (IRS) has been leveraged for improving the efficiency in wireless communication networks, which is commonly classified into three types, i.e., the fully-passive IRS, fully-active IRS, and hybrid active-passive IRS \cite{wqqtcom,10316588,passiveIRS,activeIRS,hybridIRS}. The IRS can reconfigure the signal propagation paths by altering its phase shifts, and can be exploited to enhance the sensing capability. Specifically, the IRS can create virtual LoS links between the BS and the targets, which is beneficial for scenarios where the direct links between them are blocked \cite{IRS_LoS1}. Moreover, the IRS utilizes a large aperture to enable intelligent reflection of signals, thereby countering the effects of signal attenuation and boosting the received echo signal strength for more accurate detection. Accordingly, the IRS has been investigated in wireless sensing systems \cite{kaitao,probability1,probability2,SNR1} and integrated sensing and communication (ISAC) systems \cite{ISAC_probability1,ISAC_probability2,Omni,ISAC_SNR1,ISAC_SNR2,ISAC_SNR3}. The IRS beam pattern gains were maximized for parameter estimation and target detection in \cite{kaitao}. In \cite{probability1,probability2}, the optimization of IRS reflective beamforming aimed to enhance the probability of detecting the target at a given level of false alarms. In \cite{SNR1}, IRS was deployed to control the echoes by tuning its phase shifts for enhancing the signal-to-noise ratio (SNR). In \cite{ISAC_probability1,ISAC_probability2}, a codebook design was proposed to minimize the localization error probability while meeting the users' quality-of-service requirement. \cite{Omni} investigated an intelligent omni-surface-enabled ISAC system, where the minimum SNR was maximized for sensing while accounting for communication requirements. Furthermore, the received SNR was maximized via transmission protocol design and beamforming optimization \cite{ISAC_SNR1,ISAC_SNR2,ISAC_SNR3}.
\begin{figure*}[t]
	\centering
	\subfloat[Point target case]
	{\label{fig:model_point}\includegraphics[width=0.4\textwidth]{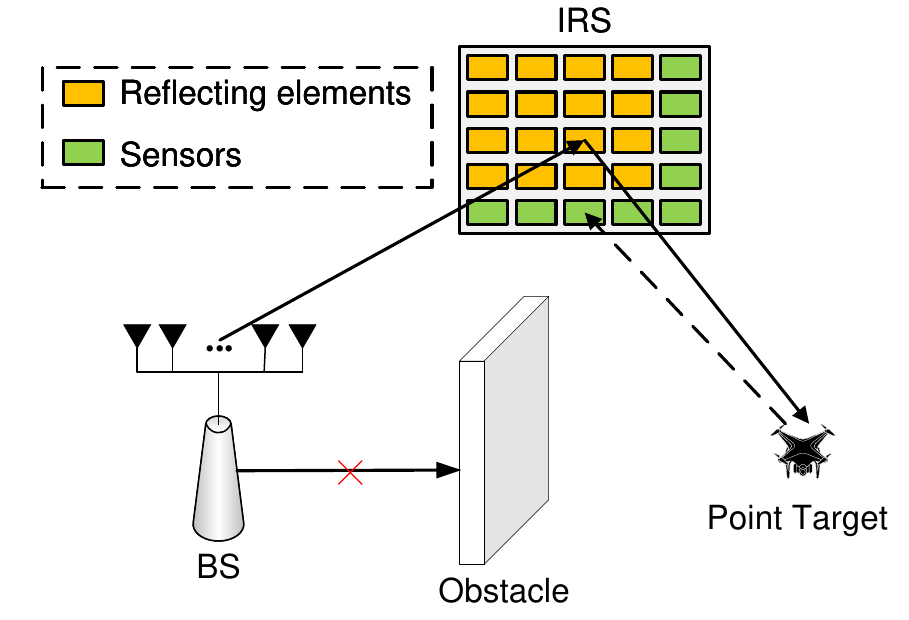}}
	\subfloat[Extended target case]{\label{fig:model_extended}\includegraphics[width=0.4\textwidth]{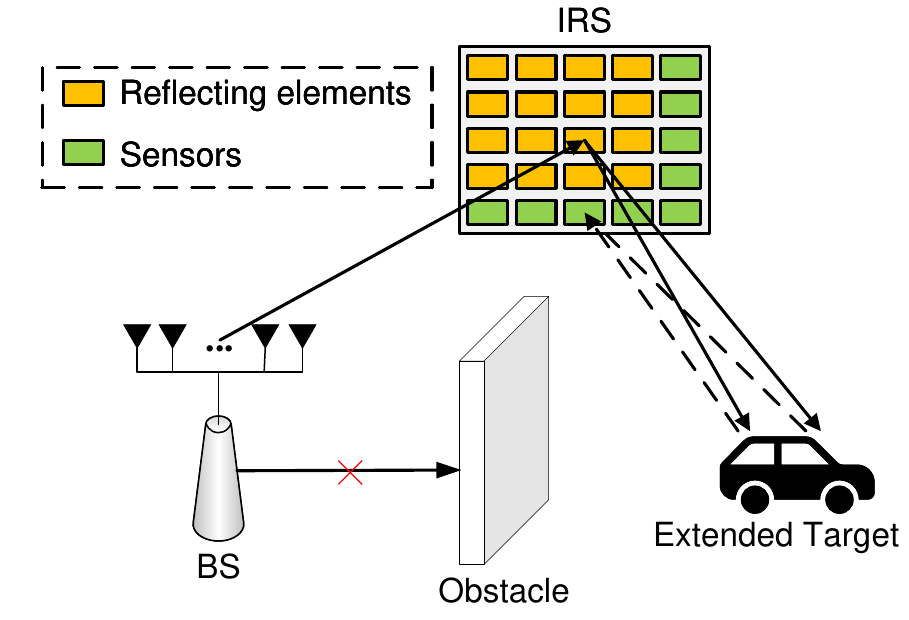}}
	\caption{A semi-passive IRS-enabled sensing system.}
	\label{fig:model}
\end{figure*}

In addition to the sensing SNR and the target detection probability, Cram\'{e}r-Rao bound (CRB) is also one of the commonly used metrics for evaluating the sensing performance, which delineates the fundamental lower bound on the covariance matrix of any unbiased estimator of the parameters. In recent works, different IRS architectures have been employed in the sensing systems for CRB minimization, i.e., the fully-passive IRS \cite{FPS,FPSISAC}, semi-passive IRS \cite{SPS1,SPS2}, and active IRS \cite{FAS}. To be specific, the fully-passive IRS only comprises a number of passive reflective elements that reflect signals to the target and echoes to the BS. In both semi-passive and active IRS-enabled sensing systems, additional active sensors are deployed as RXs with the difference being that the radar sensing signal in the latter is transmitted through the IRS controller rather than the BS in the former. For example, \cite{FPS} considered a fully-passive IRS-enabled sensing system and focused on the CRB minimization. However, its sensing performance is largely limited because the system suffers severe signal attenuation after multiple reflections, i.e., the BS-IRS-target-IRS-BS (BITIB) link. To tackle this issue, \cite{FAS} proposed an active IRS architecture and provided analytical insights on the CRB for direction-of-angle (DoA) estimation. The proposed active IRS is equipped with passive reflecting elements, dedicated active sensors, and a controller, which differs from the fully-active IRS that consists of active elements with signal amplification capabilities in communication systems. In this case, the IRS controller functions as a TX, enabling the transmission of probing signals, which makes it similar to a traditional radar station. Although this is beneficial for sensing, it incurs not only greater hardware expenses and energy usage at the IRS but also complicated and effective interference cancellation techniques, which may severely constrain the application scenarios of active IRS-enabled sensing in practice. In light of this, the semi-passive IRS architecture emerges as a cost-effective solution for wireless sensing \cite{SPS1,SPS2}. Different from the fully-passive IRS, the semi-passive IRS consists of a number of additional active sensors, which directly receive and process the target echo signals, thereby enabling the signals to travel through a less-hop link, i.e., the BS-IRS-target-sensor (BITS) link. As such, the intrinsic limitations on the sensing performance of the semi-passive IRS can be theoretically characterized by capturing the effect of the system parameters, e.g., the number of BS transmit antenna, IRS reflecting elements, and active sensors. Compared to the active IRS, the semi-passive IRS is incapable of proactively transmitting signals, thus the BS transmit beamforming needs to be jointly designed with the IRS phase shifts for CRB minimization. To this end, it remains unexplored how to design a CRB-based scheme to improve the sensing performance in such systems. 

Motivated by these considerations, this paper investigates the CRB minimization problem in a semi-passive IRS-enabled sensing system, in which an IRS is deployed for enhancing target parameters estimation as depicted in Fig. \ref{fig:model}. For a complete study, we discuss two cases in the following, i.e., the point target case and the extended target case. The objective involves minimizing the CRB for parameter estimation and analyzing the correlation between CRB and relevant system parameters under both cases. Note that the CRB characterization problem is challenging since it involves the joint design of the BS beamformer and the IRS phase shifts. The main contributions are summarized as follows:
\begin{itemize}
	\item We first study the semi-passive IRS-enabled sensing system under the point target case and obtain the corresponding CRB. The formulation of the CRB minimization problem is aimed at enhancing the accuracy of DoA estimation. To obtain some useful insights, we analyze the problem for the case with a single-antenna BS and derive the optimal solution. Furthermore, we investigate the issue of IRS element and sensor allocation for CRB minimization. Specifically, the optimal number of reflecting elements and that of sensors at the IRS for CRB minimization are derived, subject to a constraint of power/cost/total number for them. Since the formulated CRB minimization problem is more challenging for the case with a multi-antenna BS, we apply alternating optimization (AO) and semi-definite relaxation (SDR) techniques to obtain a high-quality solution. 
	\item Then, we derive the CRB for the target response matrix (TRM) estimation for the case with an extended target. Given that the CRB does not rely on the design of the reflective beamformer, the corresponding CRB minimization problem is solved by optimizing the transmit beamforming. Moreover, we obtain the closed-form expression of CRB based on our proposed design by performing singular value decomposition (SVD). The correlation between the derived CRB and system parameters is unveiled, which differs from the conclusion under the point target case. Furthermore, we compare the semi-passive IRS-enabled sensing system and the fully-passive IRS-enabled sensing system in terms of the CRB. 
	\item Numerical results illustrate the effectiveness of the proposed scheme in the semi-passive IRS-enabled sensing systems in terms of the sensing performance measured by the CRB. Moreover, our results also reveal the correlation between the CRB and various system parameters. To be specific, the CRB for the DoA estimation decreases with the transmit power, the number of BS transmit antennas, IRS reflecting elements, and IRS sensors. Furthermore, it is shown that the CRB for the TRM estimation decreases with the transmit power and the number of BS transmit antennas and increases with the number of reflecting elements and of sensors at the IRS. 
\end{itemize}

The rest of this paper is organized as follows. Section \ref{System Model} presents the system model for the semi-passive IRS-enabled sensing system. We analyze the CRB for the DoA estimation under the point target case with a single-antenna BS and a multi-antenna BS, respectively, in Section \ref{Point Target Case}. Section \ref{Extended Target Case} presents the transmit beamforming design and derives the optimal CRB for the TRM estimation in closed-form under the extended target case. Simulation results are presented in Section \ref{Numerical results} and finally, Section \ref{Conclusion} brings the conclusion.

{\it{Notations:}} Matrices and vectors are denoted by boldface upper-case and lower-case letters, respectively. $\mathbb{C}^{r_1 \times r_2}$ and $\mathbb{R}^{r_1 \times r_2}$ denote the complex-valued and real-valued matrices of dimensions $r_1 \times r_2$, respectively. For a complex-valued vector $\mathbf{s}$, $\arg \left( \mathbf{s} \right)$ represents its phase,  ${\left[\mathbf{s}\right]}_n$ denotes the $n$-th entry of $\mathbf{s}$, and $\operatorname{diag} \left(\mathbf{s}\right)$ denotes a diagonal matrix with its diagonal entries taken from the vector $\mathbf{s}$. For a square matrix $\mathbf{B}$, $\mathbf{B} \succeq \mathbf{0}$ implies that $\mathbf{B}$ is positive semi-definite and $\operatorname{tr} \left( \mathbf{B} \right)$ denotes its trace. For a general matrix $\mathbf{S}$, $\operatorname{vec} \left(\mathbf{S}\right)$, and $\operatorname{rank} \left( \mathbf{S} \right)$ represent its vector and rank, respectively. $\mathbf{I}_{R}$ is an identity matrix of dimensions $R \times R$. $\mathbf{0}$ represents an all-zero matrix. $\left( \cdot \right) ^{*}$, $\left( \cdot \right) ^T$, $\frac{{\partial}}{{\partial }} \left( \cdot \right)$, $\left( \cdot \right) ^H$ and $\|\cdot\|$ stand for the conjugate, transpose, partial derivative and conjugate transpose operators, and Euclidean norm, respectively. For a complex scalar $x$, $\left| x \right|$, $\operatorname{Im} \left( x \right)$, and $\operatorname{Re} \left( x \right)$ denote the absolute value, the imaginary part and the real part of $x$, respectively. $j$ denotes the imaginary unit, i.e., $j^2 = -1$. For a circularly symmetric complex Gaussian random variable $x$ with mean $\mu$ and variance $\sigma^2$, the notation is $x \sim \mathcal{CN} ( \mu,\sigma^2 )$. $\mathcal{O} \left( \cdot \right)$ expresses the big-O notation. $\otimes$ represents the Kronecker product.

\section{System Model}
\label{System Model}
As illustrated in Fig. \ref{fig:model}, a semi-passive IRS-enabled sensing system is considered, which includes a BS with $M$ antennas, a target, and a semi-passive IRS with $N_0 = {N_{\text{h}}} \times {N_{\text{v}}}$ reflecting elements and $K_0 = 2{K _\text{h}} - 1 = 2{K _\text{v}} -1 $ sensors.\footnote{To explore the potential advantages of semi-passive IRS in a sensing system for minimizing the CRB, we assume that there is one target. We can analyze the CRB for angle estimation and response matrix estimation similar to the one without IRS in \cite{multi-target}. Note that the AO and SVD-based approaches are applicable to the multi-target case.} The probing signals are reflected by passive reflecting elements arrayed horizontally and vertically in the number of ${N_{\text{h}}}$ and ${N_{\text{v}}}$, respectively. ${K_{\text{h}}}$ sensors and ${K_{\text{v}}}$ sensors are respectively placed horizontally and vertically, which are active and accountable for the reception and processing of the echo signal emitted by the target. The inter-element spacing and inter-sensor spacing at the semi-passive IRS are given by ${d_\text{I}}$ and ${d_\text{s}}$, respectively. For simplicity, this paper considers the estimation of the target's azimuth angle relied on the signals received by the $K _\text{h}$ horizontal sensors, and the proposed design is also applicable to its elevation angle estimation. We assume that the BS-target direct link is obstructed due to dense obstacles. 

Denote the vector of transmitted signal at symbol $t \in \mathcal{T} = \left\{ {1, \ldots ,T} \right\}$ by ${\mathbf{x}}\left( t \right) \in \mathbb{C}^{M \times 1}$. The transmit covariance matrix is given by
\begin{align}
	{{\mathbf{R}}_{\text{x}}} = \frac{1}{T}\sum\limits_{t \in \mathcal{T}} {\mathbf{x}} \left( t \right){\mathbf{x}}{\left( t \right)^H}.
\end{align}
The TRM $\mathbf{H}$ varies according to the particular target models. To be specific, we take into account the following two cases, i.e., the point and extended target cases.
 
\subsubsection{Point Target Case}
In this case, the target is an object that is small relative to the radar resolution cell and can be modeled as an unstructured point that reflects the signal from a single scatterer within the space (see Fig. \ref{fig:model} (a)). To facilitate the channel description, we denote the steering vector function for a uniform linear array (ULA) by ${\mathbf{u}}\left( {\bar \vartheta ,\bar N} \right)$, which can be expressed as
\begin{align}
	{\mathbf{u}}\left( {\bar \vartheta ,\bar N} \right) \buildrel \Delta \over = {\left[ {{e^{\frac{{ - j\left( {\bar N - 1} \right)\pi \bar \vartheta }}{2}}},{e^{\frac{{ - j\left( {\bar N - 3} \right)\pi \bar \vartheta }}{2}}}, \ldots ,{e^{\frac{{j\left( {\bar N - 1} \right)\pi \bar \vartheta }}{2}}}} \right]^T},
\end{align}
where $\bar \vartheta$ represents the steering vector direction. $\bar N$ is the number of elements of the ULA. Then, the TRM is modeled as
\begin{align}
	\label{channel}
	{\mathbf{H}} = \alpha {\mathbf{b}}\left( {{\theta _{\text{h}}}} \right){{\mathbf{a}}^T}\left( {{\theta _{\text{h}}},{\theta _{\text{v}}}} \right),
\end{align}
where the channel coefficient is denoted by $\alpha  = {\alpha _0}{\beta _0} \in \mathbb{C}$, determined by both the round-trip path-loss and the radar cross section (RCS). The small-scale fading ${\alpha _0} \sim \mathcal{CN} \left( {0,1} \right)$ refers to rapid fluctuation of the amplitude and phase of the received signal. ${\beta _0} = \sqrt {\frac{{{\lambda_\text{R}^{2}}\kappa }}{{64{\pi ^3}d_{{\text{IT}}}^4}}}$ with $\lambda_\text{R}$ denoting the carrier wavelength, $d_\text{IT}$ representing the IRS-target distance, and $\kappa$ denoting the RCS. In addition, the steering vector at the IRS sensors is expressed as
${\mathbf{b}}\left( {{\theta _{\text{h}}}} \right) = {\mathbf{u}}\left( {\frac{{2{d_\text{s}}}}{\lambda_\text{R}}\sin \left( {{\theta _{\text{h}}}} \right),{K_{\text{h}}}} \right) \in {\mathbb{C}^{{K_{\text{h}}} \times 1}}$
and the response vector function of the IRS is given by ${\mathbf{a}}\left( {{\theta _{\text{h}}},{\theta _{\text{v}}}} \right) = {\mathbf{u}}\left( {\frac{{2{d_\text{I}}}}{\lambda_\text{R}}\sin \left( {{\theta _{\text{h}}}} \right)\sin \left( {{\theta _{\text{v}}}} \right),{N_{\text{h}}}} \right) \otimes {\mathbf{u}}\left( {\frac{{2{d_\text{I}}}}{\lambda_\text{R}}\cos \left( {{\theta _{\text{v}}}} \right),{N_{\text{v}}}} \right)$, where ${{\theta _{\text{h}}}}$ and ${{\theta _{\text{v}}}}$ denote the azimuth and vertical angles-of-departure from the IRS to the target, respectively. To obtain insights, we consider the special case with a ULA of $N$ reflecting elements and $K$ sensors at the IRS.\footnote{The proposed algorithm is extendable for scenarios involving a uniform planar array IRS. This is due to the fact that the estimation of horizontal and vertical directions can be treated independently \cite{FAS}.} Assuming that $N$ and $K$ are even integers ($N \ge 2$ and $K \ge 2$), and the inter-element distance and the inter-sensor distance are the same, i.e., $d_\text{s} = d_\text{I} = \hat d$, the TRM in \eqref{channel} with the target's DoA $\theta$ is rewritten as ${\mathbf{H}} = \alpha {\mathbf{b}}( \theta ){{\mathbf{a}}^T}( \theta )$, where 
\begin{align}
	\label{a}
	{\mathbf{a}}( {{\theta}} ) \!=\!& {\left[\! {{e^{ - \frac{j \pi{( {N - 1} ){{\hat d}}}\sin ( {{\theta}} )}{\lambda_\text{R}} }},\!{e^{ - \frac{j \pi{( {N - 3} ){{\hat d}}}\sin ( {{\theta}} )}{\lambda_\text{R}} }},\! \ldots \!,\!{e^{\frac{j \pi{( {N - 1} ){{\hat d}}}\sin ( {{\theta}} )}{\lambda_\text{R}} }}}\! \right]^T}, \\
	\label{b}
	{\mathbf{b}}( \theta ) \!=\!& {\left[\! {{e^{ - \frac{j \pi{( {K - 1} ){{\hat d}}}\sin ( {{\theta}} )}{\lambda_\text{R}} }},\!{e^{ - \frac{j \pi{( {K - 3} ){{\hat d}}}\sin ( {{\theta}} )}{\lambda_\text{R}} }},\! \ldots \!,\!{e^{\frac{j \pi{( {K - 1} ){{\hat d}}}\sin ( {{\theta}} )}{\lambda_\text{R}} }}}\! \right]^T}. 
\end{align}

\subsubsection{Extended Target Case}
In this case, the size of the target exceeds the radar range resolution. The extended target can be approximated by an object composed of distributed point-like scatterers and reflects the incident signal from multiple scatterers dispersed over a larger spatial area (see Fig. \ref{fig:model} (b)). The mathematical representation of the problem for the extended target case is similar to that for the point target case if the received signal is processed individually at each sensor. Due to the impact of target spatial range and scattering characteristics on analysis and application, extended targets and point targets are not directly related in general. Denote the set of the resolvable scatterers by $\mathcal{S} = \left\{ {1, \ldots ,N_{\text{s}}} \right\}$. For the scatterer $s \in \mathcal{S}$, ${\alpha _s}$ denotes the reflection coefficient, and the angle is denoted by ${\theta _s}$. In this case, The TRM can be modeled as  
\begin{align}
	{\mathbf{H}} = \sum\limits_{s \in \mathcal{S}} {{\alpha _s} {\mathbf{b}}\left( {{\theta _s}} \right){{\mathbf{a}}^T}\left( {{\theta _s}} \right)}.
\end{align}

The IRS reflection-coefficient matrix is denoted by ${\mathbf{\Phi }} = {\operatorname{diag}} \left( {\mathbf{v}} \right)$, where ${\mathbf{v}} = {\left[ {{e^{j{\phi _{1}}}}, \ldots ,{e^{j{\phi _{N}}}}} \right]^T}$ with the phase shift ${\phi _{n}} \in \left[ {0,2\pi } \right)$, $n \in \left\{ {1, \ldots ,N} \right\}$. At symbol $t$, we can express the received echo signal as 
\begin{align}
	{\mathbf{y}}\left( t \right) = {\mathbf{H} \mathbf{\Phi} \mathbf{Gx}}\left( t \right) + {\mathbf{n}}\left( t \right),
\end{align}
where $\mathbf{G} \in \mathbb{C}^{N \times M}$ represents the channel from the BS to the semi-passive IRS, and ${\mathbf{n}}\left( t \right) \sim {\mathcal{CN}} \left( {{\mathbf{0}},\sigma _{\text{R}}^2{{\mathbf{I}}_{K}}} \right)$ represents the additive disturbance at the IRS sensors with the noise power $\sigma _{\text{R}}^2$. By stacking the received signals, we represent them in a matrix form as   
\begin{align}
	{\mathbf{Y}} = \left[ {{\mathbf{y}}\left( 1 \right),{\mathbf{y}}\left( 2 \right), \ldots, {\mathbf{y}}\left( T \right)} \right] = {\mathbf{H\Phi GX}} + {\mathbf{N}},
	\label{Y_1}
\end{align} 
where ${\mathbf{X}} = \left[ {{\mathbf{x}}\left( 1 \right),{\mathbf{x}}\left( 2 \right), \ldots, {\mathbf{x}}\left( T \right)} \right]$ and ${\mathbf{N}} = \left[ {{\mathbf{n}}\left( 1 \right),{\mathbf{n}}\left( 2 \right), \ldots, {\mathbf{n}}\left( T \right)} \right]$. Based on \eqref{Y_1}, we analyze the CRB for the DoA estimation under the point target case and for the TRM estimation under the extended target case in the following sections.
 
\section{CRB Analysis and Optimization for Point Target Case}
\label{Point Target Case}
In this section, we investigate the CRB minimization problem under the point target case. To obtain useful insights, we first derive the optimal solution for a scenario involving a single-antenna BS. Then, an efficient algorithm is developed for the corresponding problem to derive its high-quality solution for the case with a multi-antenna BS. 

In this case, we rewrite the received echoes $\mathbf{Y}$ in \eqref{Y_1} as
\begin{align}
	{\mathbf{Y}} = \alpha {\mathbf{E}}( \theta ){\mathbf{X}} + {\mathbf{N}},
	\label{Y}
\end{align}
where ${\mathbf{E}}( \theta ) = {\mathbf{b}}( \theta ){{\mathbf{a}}^T}( \theta ){\mathbf{\Phi G}}$. For ease of notation, we denote ${\mathbf{a}}( \theta )$, ${\mathbf{b}}( \theta )$, and ${\mathbf{E}}( \theta )$ as ${\mathbf{a}}$, ${\mathbf{b}}$ and ${\mathbf{E}}$, respectively. To facilitate the CRB derivation, we convert ${\mathbf{Y}}$ in \eqref{Y} into a vector form as
\begin{align}
\label{y}
{\mathbf{\hat y}} = {\operatorname{vec}} \left( {\mathbf{Y}} \right) = {\mathbf{\hat z}} + {\mathbf{\hat n}},
\end{align}
where ${\mathbf{\hat z}} = \alpha {\operatorname{vec}} \left( {{\mathbf{EX}}} \right)$. ${\mathbf{\hat n}}$ is the vectorization of the matrix ${\mathbf{N}}$, and is distributed as $\mathcal{CN} ( {{\mathbf{0}},{{\mathbf{\hat R}}_{\text{n}}}} )$ with ${{\mathbf{\hat R}}_{\text{n}}}=\sigma _{\text{R}}^2{{\mathbf{I}}_{KT}}$. The unknown parameters are defined by ${\bm{\xi}} = {\left[ {\theta ,{{{\bm{\hat \alpha }}}^T}} \right]^T} \in {\mathbb{R}}{^{3 \times 1}}$ with ${\bm{\hat \alpha }} = {\left[ {{\operatorname{Re}} \left\{ \alpha  \right\},{\operatorname{Im}} \left\{ \alpha  \right\}} \right]^T}$. To begin with, we derive the Fisher information matrix (FIM) ${{\mathbf{\hat F}}} \in {\mathbb{R}}{^{3 \times 3}}$ for estimating ${\bm{\xi}}$, where the $(p,q)$ entry of ${{\mathbf{\hat F}}}$ is given by \cite{kay1993fundamentals}
\begin{align}
	\label{F_point}
	{[ {{\mathbf{\hat F}}} ]_{p,q}} =& \operatorname{tr} \left( {{\mathbf{\hat R}}_{\text{n}}^{ - 1}\frac{{\partial {{\mathbf{\hat R}}_{\text{n}}}}}{{\partial {{\bm{\xi }}_p}}}{\mathbf{\hat R}}_{\text{n}}^{ - 1}\frac{{\partial {{\mathbf{\hat R}}_{\text{n}}}}}{{\partial {{\bm{\xi }}_q}}}} \right) \nonumber\\
	&+ 2 \operatorname{Re} \left\{ {\frac{{\partial {{{\mathbf{\hat z}}}^H}}}{{\partial {{\bm{\xi }}_p}}}{\mathbf{\hat R}}_{\text{n}}^{ - 1}\frac{{\partial {\mathbf{\hat z}}}}{{\partial {{\bm{\xi }}_q}}}} \right\},p,q \in \left\{ {1,2,3} \right\}.
\end{align}
Based on \eqref{F_point}, we write the FIM in a block matrix form, which is given by
\begin{align}
\label{FIM_point}
{\mathbf{\hat F}} = \left[ {\begin{array}{*{20}{c}}
{{{{\hat F}}_{\theta \theta }}}&{{{\mathbf{\hat F}}_{\theta {\bm{\hat \alpha }}}}}\\
{{\mathbf{\hat F}}_{\theta {\bm{\hat \alpha }}}^T}&{{{\mathbf{\hat F}}_{{\bm{\hat \alpha \hat \alpha }}}}}
\end{array}} \right],
\end{align}
where
\begin{align}
&{{{\hat F}}_{\theta \theta }} = \frac{{2T{{\left| \alpha  \right|}^2}}}{{\sigma _{\text{R}}^2}}{\operatorname{tr}} \left( {{\mathbf{\dot E}}{{\mathbf{R}}_{\text{x}}}{{{\mathbf{\dot E}}}^H}} \right), \label{FIM_point_1}\\
&{{\mathbf{\hat F}}_{\theta {\bm{\hat \alpha }}}} = \frac{{2T}}{{\sigma _{\text{R}}^2}}{\operatorname{Re}} \left\{ {{\alpha ^*}{\operatorname{tr}} \left( {{\mathbf{E}}{{\mathbf{R}}_{\text{x}}}{{{\mathbf{\dot E}}}^H}} \right)\left[ {1,j} \right]} \right\}, \label{FIM_point_2}\\
&{{\mathbf{\hat F}}_{{\bm{\hat \alpha \hat \alpha }}}} = \frac{{2T}}{{\sigma _{\text{R}}^2}}{\operatorname{tr}} \left( {{\mathbf{E}}{{\mathbf{R}}_{\text{x}}}{{\mathbf{E}}^H}} \right){{\mathbf{I}}_2},\label{FIM_point_3}
\end{align}
where ${\mathbf{\dot E}} = \frac{{\partial {\mathbf{E}}}}{{\partial \theta }}$. The details of the above derivations can be found in Appendix A. We focus on the estimation of $\theta$. Based on \eqref{FIM_point}, \eqref{FIM_point_1}, \eqref{FIM_point_2} and \eqref{FIM_point_3}, the corresponding CRB matrix is given by
\begin{align}
\label{crb}
{\text{CRB}}( \theta ) &= [{{\mathbf{\hat F}}^{ - 1}}]_{1,1} \nonumber\\
&= {\left[ {{{{\hat F}}_{\theta \theta }} - {{\mathbf{\hat F}}_{\theta {\bm{\hat \alpha }}}}{{\mathbf{\hat F}}_{{\bm{\hat \alpha \hat \alpha }}}^{-1}}{\mathbf{\hat F}}_{\theta {\bm{\hat \alpha }}}^T} \right]^{ - 1}} \nonumber\\
&= \frac{{\sigma _{\text{R}}^2}}{{2T{{\left| \alpha  \right|}^2}\left( {{\operatorname{tr}} \left( {{\mathbf{\dot E}}{{\mathbf{R}}_{\text{x}}}{{{\mathbf{\dot E}}}^H}} \right) - \frac{{{{\left| {{\operatorname{tr}} \left( {{\mathbf{E}}{{\mathbf{R}}_{\text{x}}}{{{\mathbf{\dot E}}}^H}} \right)} \right|}^2}}}{{{\operatorname{tr}} \left( {{\mathbf{E}}{{\mathbf{R}}_{\text{x}}}{{\mathbf{E}}^H}} \right)}}} \right)}}.
\end{align}
Since we choose the array centroid as the reference point in \eqref{a} and \eqref{b}, we have that due to the symmetry,
\begin{align}
	\label{symmetry}
	{{\mathbf{\dot a}}^H}{\mathbf{a}} = 0,{{\mathbf{a}}^H}{\mathbf{\dot a}} = 0,
	{{\mathbf{\dot b}}^H}{\mathbf{b}} = 0,{{\mathbf{b}}^H}{\mathbf{\dot b}} = 0,\forall \theta,
\end{align}
where the derivative of $\mathbf{a}$ and $\mathbf{b}$ are given by 
\begin{align}
	{\mathbf{\dot a}} &= \frac{{\partial {\mathbf{a}}}}{{\partial \theta }}= j \pi \frac{{{\hat d}}}{{{\lambda_\text{R}}}}\cos ( \theta ){{\mathbf{D}}_{\text{a}}}{\mathbf{a}}, \label{adot}\\
	{\mathbf{\dot b}} &= \frac{{\partial {\mathbf{b}}}}{{\partial \theta }}= j \pi \frac{{{\hat d}}}{{{\lambda_\text{R}}}}\cos ( \theta ){{\mathbf{D}}_{\text{b}}}{\mathbf{b}}, \label{bdot}
\end{align}
with ${{\mathbf{D}}_{\text{a}}} = {\operatorname{diag}} ( { - ( {N - 1} ), - ( {N - 3} ), \ldots , ( {N - 1} )} )$ and ${{\mathbf{D}}_{\text{b}}} = {\operatorname{diag}} ( { - ( {K - 1} ), - ( {K - 3} ), \ldots , ( {K - 1} )} )$. By introducing ${\mathbf{A}} = {\operatorname{diag}} \left( {{\mathbf{a}}} \right)$, we have 
\begin{align}
	{\mathbf{E}} &= {\mathbf{b}}{{\mathbf{a}}^T}{\mathbf{\Phi G}} = {\mathbf{b}}{{\mathbf{v}}^T}{\mathbf{A G}}, \label{BBB}\\
	\mathbf{\dot E} &= {{ {{\mathbf{\dot b}}{{\mathbf{a}}^T}{\mathbf{\Phi G}} + {\mathbf{b}}{{{\mathbf{\dot a}}}^T}{\mathbf{\Phi G}}} }} \nonumber\\
	&= j \pi \frac{{{\hat d}}}{{{\lambda_\text{R}}}}\cos ( \theta ) \left({{\mathbf{D}}_{\text{b}}}{\mathbf{b}}{{\mathbf{v}}^T}{\mathbf{A G}} + \mathbf{b}{{\mathbf{v}}^T}{\mathbf{D}_\text{a}}{\mathbf{A G}}\right). \label{BdotBdotBdot}
\end{align}
Based on \eqref{adot}, \eqref{bdot}, \eqref{BBB} and \eqref{BdotBdotBdot} and the orthogonality property \eqref{symmetry}, it yields
\begin{align}
	{\operatorname{tr}} \left( {{\mathbf{E}}{{\mathbf{R}}_{\text{x}}}{{\mathbf{E}}^H}} \right) = & {\operatorname{tr}} \left( {\left( {{\mathbf{b}}{{\mathbf{a}}^T}{\mathbf{\Phi G}}} \right){{\mathbf{R}}_{\text{x}}}{{\left( {{\mathbf{b}}{{\mathbf{a}}^T}{\mathbf{\Phi G}}} \right)}^H}} \right) \nonumber\\
	= & {\left\| {\mathbf{b}} \right\|^2} {{{\mathbf{v}}^H}{{\mathbf{R}}_1}{\mathbf{v}}}, \label{BB} \\
	{\operatorname{tr}} \left( {{\mathbf{E}}{{\mathbf{R}}_{\text{x}}}{{{\mathbf{\dot E}}}^H}} \right) = & {\operatorname{tr}} \left( {( {{\mathbf{b}}{{\mathbf{a}}^T}{\mathbf{\Phi G}}} ){{\mathbf{R}}_{\text{x}}}{{( {{\mathbf{\dot b}}{{\mathbf{a}}^T}{\mathbf{\Phi G}} + {\mathbf{b}}{{{\mathbf{\dot a}}}^T}{\mathbf{\Phi G}}} )}^H}} \right) \nonumber\\
	= & -j \pi \frac{{{\hat d}}}{{{\lambda _{\text{R}}}}}\cos ( \theta ){\left\| {\mathbf{b}} \right\|^2} {{{\mathbf{v}}^H}{\mathbf{D}_\text{a}}{{\mathbf{R}}_1}{\mathbf{v}}}, \label{BBdot} \\
	{\operatorname{tr}} \left( {{\mathbf{\dot E}}{{\mathbf{R}}_{\text{x}}}{{{\mathbf{\dot E}}}^H}} \right) =& {\operatorname{tr}} \left( {{\mathbf{\dot b}}{{\mathbf{a}}^T}{\mathbf{\Phi G}}{{\mathbf{R}}_{\text{x}}}{{( {{\mathbf{\dot b}}{{\mathbf{a}}^T}{\mathbf{\Phi G}}} )}^H}} \right) \nonumber\\
	&+ {\operatorname{tr}} \left( {{\mathbf{b}}{{{\mathbf{\dot a}}}^T}{\mathbf{\Phi G}}{{\mathbf{R}}_{\text{x}}}{{( {{\mathbf{b}}{{{\mathbf{\dot a}}}^T}{\mathbf{\Phi G}}} )}^H}} \right)\nonumber\\
	=&  {\pi ^2}\frac{{{\hat d}^2}}{{\lambda _{\text{R}}^2}}{\cos ^2}( \theta ){\left\| {{{\mathbf{D}}_{\text{b}}}{\mathbf{b}}} \right\|^2} {{{\mathbf{v}}^H}{{\mathbf{R}}_1}{\mathbf{v}}} \nonumber\\
	&+ {\pi ^2}\frac{{{\hat d}^2}}{{\lambda _{\text{R}}^2}}{\cos ^2}( \theta ) {\left\| {\mathbf{b}} \right\|^2} {{{\mathbf{v}}^H}{\mathbf{D}_\text{a}}{{\mathbf{R}}_1}{\mathbf{D}_\text{a}}{\mathbf{v}}}, \label{BdotBdot}
\end{align}
where ${{\mathbf{R}}_1} = {{\mathbf{A}^*}}{\mathbf{G}^*}{{\mathbf{R}}^T_{\text{x}}}{{\mathbf{G}}^T}{\mathbf{A}}$.
As such, we derive the following proposition regarding the CRB and proceed to conduct theoretical analysis on it.

\begin{figure*}[t]
	\begin{align}
		\label{CRB_point}
		{\text{CRB}}( \theta ) = \frac{{\sigma _{\text{R}}^2\lambda _{\text{R}}^2}}{{2T{{\left| \alpha  \right|}^2}{\pi ^2}{{\hat d}^2}{{\cos }^2}( \theta )\left( {( {\frac{{{K^3} - K}}{3}} ) ( {{{\mathbf{v}}^H}{{\mathbf{R}}_1}{\mathbf{v}}} ) + K ( {{{\mathbf{v}}^H}{{\mathbf{D}}_{\text{a}}}{{\mathbf{R}}_1}{{\mathbf{D}}_{\text{a}}}{\mathbf{v}}} ) - K\frac{{{{\left| {{{\mathbf{v}}^H}{\mathbf{D}_\text{a}}{{\mathbf{R}}_1}{\mathbf{v}}} \right|}^2}}}{{( {{{\mathbf{v}}^H}{{\mathbf{R}}_1}{\mathbf{v}}} )}}} \right)}}.
	\end{align}
	{\noindent} \rule[0pt]{18cm}{0.05em}
\end{figure*}

\begin{Proposition}
\label{Pro_point}
	Under the point target case, the CRB for the DoA estimation is expressed in \eqref{CRB_point} on the top of this page.
\end{Proposition}

{\it{Proof:}}
Based on \eqref{b}, we have ${{\left\| {{{\mathbf{D}}_{\text{b}}}{\mathbf{b}}} \right\|}^2} = (K^3 - K)/3$ and ${{\left\| {\mathbf{b}} \right\|}^2} = K$. By substituting \eqref{BB}, \eqref{BBdot} and \eqref{BdotBdot} into \eqref{crb}, \eqref{CRB_point} is naturally obtained, which completes the proof. ~$\hfill\blacksquare$

Proposition \ref{Pro_point} shows that the CRB of $\theta$ monotonically decreases as $T$ increases. This is because more measurements provide better sampling and more diverse statistical information on the parameters being estimated, allowing the estimator to capture more variation in the data, thus improving the accuracy. Moreover, it is observed that the CRB increases with $\sigma_\text{R}^2$, because higher noise power introduces more uncertainty and randomness into the estimated data, making accurate estimation more challenging and resulting in greater variance in the estimator. Next, we further provide an analysis of the deployment of the IRS sensors.
\begin{Proposition}
	\label{CRB_K}
	Under the point target case, the CRB of $\theta$ monotonically decreases with $K$.
\end{Proposition}

{\it{Proof:}}
Note that ${\text{CRB}}\left( \theta \right)$ is dependent on $K$, and thus, we use ${\text{CRB}_\theta }\left( K \right)$ to represent ${\text{CRB}} \left( \theta \right)$. By relaxing $K$ to a continuous variable, the partial derivative of ${\text{CRB}_\theta }\left( K \right)$ with respect to (w.r.t.) $K$ is given by 
\begin{align}
	\frac{{\partial {{\text{CRB}_\theta }( K )} }}{{\partial K}} = \frac{{ - {\beta _1}( {3{\beta _2}{K^2} - {\beta _3}})}}{{{{( {{\beta _2}{K^3} - {\beta _3}K} )}^2}}},
\end{align}
where ${{\beta _1}} = {{\sigma _{\text{R}}^2\lambda _{\text{R}}^2}}/{({2T{{\left| \alpha  \right|}^2}{\pi ^2}{{\hat d}^2}{{\cos }^2}( \theta  )})}$, ${{\beta _2}} = {{{{{\mathbf{v}}^H}{{\mathbf{R}}_1}{\mathbf{v}}}}/{3}}$ and ${{\beta _3}} = {{\beta _2}}-{{{\mathbf{v}}^H}{{\mathbf{D}}_{\text{a}}}{{\mathbf{R}}_1}{{\mathbf{D}}_{\text{a}}}{\mathbf{v}}+{{{{\left|{{{\mathbf{v}}^H}{\mathbf{D}_\text{a}}{{\mathbf{R}}_1}{\mathbf{v}}} \right|}^2}}}/{{({{\mathbf{v}}^H}{{\mathbf{R}}_1}{\mathbf{v}})}}}$. Without loss of generality, ${{\text{CRB}_\theta }( K )} \textgreater 0$, i.e., ${{\beta _2}{K^2} - {\beta _3}} \textgreater0$, we have $\frac{{d\left( {{\text{CRB}_\theta }( K )} \right)}}{{dK}} \textless 0$, which completes the proof. ~$\hfill\blacksquare$

Proposition \ref{CRB_K} shows that more sensors should be deployed at the IRS to enhance the accuracy for estimating the point target's $\theta$. The result is expected because increasing the number of IRS sensors enhances the spatial resolution of sensing through the acquisition of more diverse spatial information. Moreover, it contributes to a higher array gain, which can effectively enhance the strength of the received signal and improve the extraction of estimated information from it. In contrast, increasing $K$ leads to a higher CRB for the TRM estimation under the extended target case, which will be explained in Section \ref{Extended Target Case}. 

Based on \eqref{CRB_point}, we minimize the CRB of $\theta$ via the joint optimization of the transmit and reflective beamforming. Then, the corresponding problem for CRB minimization can be formulated as
\begin{subequations}
	\label{P_original}
	\begin{align}
		\mathop {\min }\limits_{{{\mathbf{R}}_{\text{x}}},{\mathbf{v}}} \;\;\;\; &{\text{CRB}}( \theta )\\
		{\text{s.t.}}\;\;\;\;
		&{\operatorname{tr}} \left( {{{\mathbf{R}}_{\text{x}}}} \right) \le {P_0}, \label{constraint_power}\\
		&{{\mathbf{R}}_{\text{x}}} \succeq {\mathbf{0}}, \label{constraint_nonnegative}\\
		&\left| {{{\mathbf{v}}_n}} \right| = 1,\forall n \in \left\{ {1, \ldots ,N} \right\} \label{constraint_v},
	\end{align}
\end{subequations}
where ${{{\mathbf{v}}_n}}$ denotes the $n$-th element of $\mathbf{v}$, and ${P_0}$ in \eqref{constraint_power} is the BS's transmit power budget. For problem \eqref{P_original}, constraints \eqref{constraint_nonnegative} and \eqref{constraint_v} are the non-negative constraint on the optimization variable and the unit-modulus constraint for the IRS phase shift, respectively. Problem \eqref{P_original} is intractable due to the complex interdependence of the optimization variables ${\mathbf{R}}_{\text{x}}$ and ${\mathbf{v}}$ in the objective function. To address these challenges, we first solve problem \eqref{P_original} when dealing with a single-antenna BS, and then an efficient algorithm is developed to obtain high-quality solutions for a scenario involving a multi-antenna BS. 

\subsection{Single-Antenna BS}
To draw essential insights, we consider the simple system setup in which the BS is equipped with a single antenna, i.e., $M = 1$. Let $\mathbf{h}_\text{BI}$ denote the BS-IRS channel with the corresponding distance-dependent path-loss factor ${h_{{\text{BI}}}}$. As such, \eqref{BB}, \eqref{BBdot} and \eqref{BdotBdot} can be rewritten as
\begin{align}
	{\operatorname{tr}} \left( {{\mathbf{E}}{{\mathbf{R}}_{\text{x}}}{{\mathbf{E}}^H}} \right) = & {p_\text{x}}{\left\| {\mathbf{b}} \right\|^2}{\left| {{{\mathbf{a}}^T}{\mathbf{\Phi} \mathbf{h}_\text{BI}}} \right|^2}, \label{bb}
\end{align}
\begin{align}
	{\operatorname{tr}} \left( {{\mathbf{E}}{{\mathbf{R}}_{\text{x}}}{{{\mathbf{\dot E}}}^H}} \right) = & {p_\text{x}}{\left\| {\mathbf{b}} \right\|^2}{{\mathbf{a}}^T}{\mathbf{\Phi } \mathbf{h}_\text{BI}}{{\mathbf{h}}^H_\text{BI}}{{\mathbf{\Phi }}^H}{\mathbf{\dot a}^{*}},\label{bbdot}
\end{align}
\begin{align}
{\operatorname{tr}} \left( {{\mathbf{\dot E}}{{\mathbf{R}}_{\text{x}}}{{{\mathbf{\dot E}}}^H}} \right) 
= & {\pi ^2}\frac{{{\hat d}^2}}{{{\lambda_\text{R}^2}}}{\cos ^2}( \theta ){p_\text{x}}{\left\| {{{\mathbf{D}}_{\text{b}}}{\mathbf{b}}} \right\|^2}{\left| {{{\mathbf{a}}^T}{\mathbf{\Phi} \mathbf{h}_\text{BI}}} \right|^2} \nonumber\\
& + {p_\text{x}}{\left\| {\mathbf{b}} \right\|^2}{\left| {{{{\mathbf{\dot a}}}^T}{\mathbf{\Phi} \mathbf{h}_\text{BI}}} \right|^2},
\label{bdotbdot}
\end{align}
where $\mathbf{E} = {\mathbf{b}}{{\mathbf{a}}^T}{\mathbf{\Phi}\mathbf{h}_\text{BI}}$. By substituting \eqref{bb}, \eqref{bbdot} and \eqref{bdotbdot} into \eqref{crb}, the CRB of $\theta$ can be calculated as
\begin{align}
	\label{CRB_single_original}
	{\text{CRB}} ( \theta ) 
	&=\! \frac{{\sigma _{\text{R}}^2{\lambda_\text{R}^2}}}{{2T{{\left| \alpha  \right|}^2}{\pi ^2}{{\cos }^2}( \theta ){\hat d}^2 {p_\text{x}}{{\left\| {{{\mathbf{D}}_{\text{b}}}{\mathbf{b}}} \right\|}^2}{\left| {{{\mathbf{a}}^T}{\mathbf{\Phi} \mathbf{h}_\text{BI}}} \right|^2}}} \nonumber\\
	&=\! \frac{{3\sigma _{\text{R}}^2\lambda _{\text{R}}^2}}{{2T{{\left| \alpha  \right|}^2}{\pi ^2}{{\cos }^2}( \theta ){\hat d}^2 {p_\text{x}} ( {{K^3} - K} ){\left| {{{\mathbf{a}}^T}{\mathbf{\Phi} \mathbf{h}_\text{BI}}} \right|^2}}}.
\end{align}
Then, problem \eqref{P_original} can be equivalently transformed into
\begin{subequations} \label{SISO}
\begin{align}
\mathop {\max }\limits_{{p_\text{x}},{\mathbf{\Phi }}} \;\;\;\; & {p_\text{x}}{\left| {{{\mathbf{a}}^T}{\mathbf{\Phi} \mathbf{h}_\text{BI}}} \right|^2}\\
{\text{s.t.}}\;\;\;\;
&{p_\text{x}} \le {P_0},\\
&0 \le {\phi _n} \le 2 \pi,\forall n \in \left\{ {1, \ldots ,N} \right\}.
\end{align}
\end{subequations}
It can be readily verified that the optimal solutions to problem \eqref{SISO} are ${p_\text{x}^*} = P_0$ and ${\phi _n^*} = - \arg \left\{ {{{\left[ {\mathbf{a}} \right]}_n}} \right\} - \arg \left\{ {{{\left[ {{{\mathbf{h}}_{{\text{BI}}}}} \right]}_n}} \right\},\forall n \in \left\{ {1, \ldots ,N} \right\}$, which are used in the following proposition to obtain the closed-form CRB. 
\begin{Proposition} 
\label{CRB_single}
Under the point target case with a single-antenna BS, the CRB for DoA estimation can be obtained in closed-form, which is expressed as
\begin{align}
	\label{CRB_single_opt}
	{{\text{CRB}}( \theta )}^{\text{opt}} = \frac{{3\sigma _{\text{R}}^2{\lambda_\text{R}^2}}}{{2T{{\left| \alpha  \right|}^2}{\pi ^2}{{\cos }^2}( \theta ){{\hat d}^2}{h_{{\text{BI}}}^2}{P_0}{N^2}{\left( {{K^3} - K} \right)}}}.
\end{align}
\end{Proposition}

{\it{Proof:}} Substituting the optimal solution to problem \eqref{SISO} into \eqref{CRB_single_original}, the CRB of $\theta$ can be calculated as 
\begin{align}
	{{\text{CRB}}( \theta )}^{\text{opt}} &\!= \frac{{3\sigma _{\text{R}}^2{\lambda_\text{R}^2}}}{{2T{{\left| \alpha  \right|}^2}{\pi ^2}{{\cos }^2}( \theta ){{\hat d}^2}{r_{\text{x}}}{( {{K^3} - K} )}{\left| {{{\mathbf{a}}^T}{\mathbf{\Phi } \mathbf{h}_\text{BI}}} \right|^2}}} \nonumber\\
	&\!\mathop  = \limits^{(a1)} \frac{{3\sigma _{\text{R}}^2{\lambda_\text{R}^2}}}{{2T{{\left| \alpha  \right|}^2}{\pi ^2}{{\cos }^2}( \theta ){{\hat d}^2}{h_{{\text{BI}}}^2}{r_{\text{x}}}{N^2}{( {{K^3} - K} )}}} \nonumber\\
	&\!\mathop  = \limits^{(a2)} \frac{{3\sigma _{\text{R}}^2{\lambda_\text{R}^2}}}{{2T{{\left| \alpha  \right|}^2}{\pi ^2}{{\cos }^2}( \theta ){{\hat d}^2}{h_{{\text{BI}}}^2}{P_0}{N^2}{( {{K^3} - K} )}}},
\end{align}
where $\left(a1\right)$ utilizes the optimal design of ${\mathbf{\Phi }}$, i.e., ${\phi _n^{*}} = - \arg \left\{ {{{\left[ {\mathbf{a}} \right]}_n}} \right\} - \arg \left\{ {{{\left[ {{{\mathbf{h}}_{{\text{BI}}}}} \right]}_n}} \right\},\forall n \in \left\{ {1, \ldots ,N} \right\}$ and $\left(a2\right)$ utilizes $ {p_{\text{x}}^*}={P_0}$. ~$\hfill\blacksquare$

Proposition \ref{CRB_single} shows that the CRB decreases inversely proportional to $N^2 (K^3 - K)$ since it is influenced by both the received power and the phase difference among the sensors at the IRS. To be specific, the IRS elements provide the squared-power passive beamforming gain \cite{N2}, whereas the IRS sensors contribute the total gain of $\mathcal{O}(K^3)$, including the receive beamforming gain of $\mathcal{O}(K)$ and the spatial direction gain of $\mathcal{O}(K^2)$. 

Inspired by the above analysis in Proposition \ref{CRB_single}, we further investigate the optimal configuration of system parameters for CRB minimization. Let ${Q_{{\text{tot}}}} = Q_{\text{I}} + Q_{\text{s}}$ denote the constraint of power/cost/total number for the number of IRS elements and sensors. By relaxing the value of $N$ and $K$ into continuous values ${\tilde N}$ and ${\tilde K}$, we have $Q_{\text{I}} = \tilde N{W_{\text{I}}}$ and $Q_{\text{s}} = \tilde K{W_{\text{s}}}$, where ${W_{\text{I}}}$ and ${W_{\text{s}}}$ are the corresponding weights. With the constraint, we derive the following proposition for the IRS element allocation design.

\begin{Proposition}
\label{pro_NK_opt}
The optimal number of the IRS passive reflecting elements and that of the active sensors for CRB minimization are respectively given by
\begin{align}
	\label{opt_NK}
	\tilde{N}^\text{opt} &= {{{ {Q_{{\text{tot}}}}}}/({{( {1 + \varsigma_1 } ){W_{\text{I}}}}})}, \nonumber\\
	\tilde{K}^\text{opt} &= {{{\varsigma_1 {Q_{{\text{tot}}}}}}/({{( {1 + \varsigma_1 } ){W_{\text{s}}}}})}, 
\end{align}
where $\varsigma_1  \!=\! - 1/{\beta _4}  + \sqrt[3]{{( {\beta _6} + {\beta _7})/(2\beta _4^3)}} + \sqrt[3]{{( {\beta _6} - {\beta _7})/(2\beta _4^3)}}  $, ${\beta _4} \!=  - 2{{( {Q_{{\text{tot}}}^2 - W_{\text{s}}^2} )}}/{{( {Q_{{\text{tot}}}^2 + W_{\text{s}}^2} )}} $, ${\beta _5} = - {{W_{\text{s}}^2}}/{{( {Q_{{\text{tot}}}^2 + W_{\text{s}}^2} )}}$, ${\beta _6} = -\beta _4^2{\beta _5}-2$ and $\beta _7 = \sqrt{\beta _6^2 - 4}$.
To obtain useful insights, we also provide a high-quality solution, which is expressed as
\begin{align}
	\label{subopt_NK}
	\tilde{N}^\text{sub-opt} &= \frac{{2Q_{{\text{tot}}}^3 - 2{Q_{{\text{tot}}}}W_{\text{s}}^2}}{{5Q_{{\text{tot}}}^2W_{\text{I}} + W_{\text{s}}^2}W_{\text{I}}} ,\nonumber\\
	\tilde{K}^\text{sub-opt} &= \frac{{3Q_{{\text{tot}}}^3 + 3{Q_{{\text{tot}}}}W_{\text{s}}^2}}{{5Q_{{\text{tot}}}^2{W_{\text{s}}} + W_{\text{s}}^3}}.
\end{align} 

\end{Proposition}
{\it{Proof:}} Please refer to Appendix C. ~$\hfill\blacksquare$

Proposition \ref{pro_NK_opt} provides helpful guidance for the IRS element and sensor allocation design. One can observe from \eqref{subopt_NK} that the number of IRS elements exceeds that of sensors if $W_{\text{I}} \in (0,  ({{2Q_{{\text{tot}}}^2W_{\text{s}} - 2W_{\text{s}}^3}})/({{3Q_{{\text{tot}}}^2 + 3W_{\text{s}}^2}}))$. However, the condition is not satisfied when $W_{\text{I}}=W_{\text{s}}=1$ for any $Q_{{\text{tot}}}$. As such, under the case with the constraint of total number, more sensors should be deployed at the semi-passive IRS. This is expected because equipping the IRS with a greater number of sensors under the same weight configuration will result in a higher overall gain. Our findings will be validated via simulations in Section \ref{Numerical results}.

\subsection{Multi-Antenna BS}
Before solving problem \eqref{P_original} under the case with a multi-antenna BS, it can be equivalently converted into a form that is more tractable. We define $f \left( {{{\mathbf{R}}_{\text{x}}, \mathbf{V}}} \right) = \frac{{{K^2} - 1}}{3}{\operatorname{tr}}({{\mathbf{A}}^*}{{\mathbf{G}}^*}{\mathbf{R}}_{\text{x}}^T{{\mathbf{G}}^T}{\mathbf{AV}}) + {\operatorname{tr}}({{\mathbf{D}}_{\text{a}}}{{\mathbf{A}}^*}{{\mathbf{G}}^*}{\mathbf{R}}_{\text{x}}^T{{\mathbf{G}}^T}{\mathbf{A}}{{\mathbf{D}}_{\text{a}}}{\mathbf{V}}) - {{{{\left| {{\operatorname{tr}}({{\mathbf{A}}^*}{{\mathbf{G}}^*}{\mathbf{R}}_{\text{x}}^T{{\mathbf{G}}^T}{\mathbf{A}}{{\mathbf{D}}_{\text{a}}}{\mathbf{V}})} \right|}^2}}}/{{{\operatorname{tr}}({{\mathbf{A}}^*}{{\mathbf{G}}^*}{\mathbf{R}}_{\text{x}}^T{{\mathbf{G}}^T}{\mathbf{AV}})}}$, where the semi-definite matrix $\mathbf{V} = {\mathbf{v}}{\mathbf{v}^H}$ satisfying $\operatorname{rank}(\mathbf{V}) = 1$. Without the non-convex rank-one constraint, the SDR reformulation of problem \eqref{P_original} is expressed as
\begin{subequations}
	\label{P_new}
	\begin{align}
		\mathop {\max }\limits_{{{\mathbf{R}}_{\text{x}}, \mathbf{V}}} \;\;\;\; &f \left( {{{\mathbf{R}}_{\text{x}}, \mathbf{V}}} \right) \label{f_obj} \\
		\text{s.t.} \;\;\;\; &{{\mathbf{V}}_{n,n}} = 1,\forall n = 1, \ldots ,N, \label{constraint_rankV}\\
		&\mathbf{V} \succeq \mathbf{0} \label{constraint_V}, \\
		&\eqref{constraint_power}, \eqref{constraint_nonnegative}.
	\end{align}
\end{subequations}
Solving problem \eqref{P_new} optimally is challenging because of the coupling of the optimization variables $\mathbf{R}_\text{x}$ and $\mathbf{V}$ in \eqref{f_obj}, which causes the non-convexity of problem \eqref{P_new}. To address this difficulty, the AO method is applied to iteratively optimize $\mathbf{R}_\text{x}$ and $\mathbf{V}$ until convergence. Specifically, with fixed either $\mathbf{R}_\text{x}$ or $\mathbf{V}$, problem \eqref{P_new} is reduced to a typical semi-definite program (SDP), which can be solved by CVX directly. 

\subsubsection{Transmit Beamforming Optimization}
For any fixed $\mathbf{V}$, define $\mathbf{R}_2 = \frac{{{K^2} - 1}}{3}{{\mathbf{G}}^H}{{\mathbf{A}}^H}{{\mathbf{V}}^T}{\mathbf{AG}} + {{\mathbf{G}}^H}{{\mathbf{A}}^H}{{\mathbf{D}}_{\text{a}}}{{\mathbf{V}}^T}{{\mathbf{D}}_{\text{a}}}{\mathbf{AG}}$ and $\mathbf{J} = {{\mathbf{G}}^H}{{\mathbf{A}}^H}{{\mathbf{V}}^T}$. By adopting the Schur's complement condition with the introduced auxiliary variable $t_2$, problem \eqref{P_new} is reformulated as
\begin{subequations}
	\label{TBO}
	\begin{align}
			\mathop {\max }\limits_{{{\mathbf{R}}_{\text{x}},t_1}} \;\;\;\; & t_1\\
			\text{s.t.} \;\;\;\;\; &\!\!\!\!\!\! \! \left[ {\begin{array}{*{20}{c}}
							{ {\operatorname{tr}}({\mathbf{R}_2}{{\mathbf{R}}_{\text{x}}}) - {t_1}}& \!\!\! {{\operatorname{tr}}({{\mathbf{J}}}{{\mathbf{D}}_{\text{a}}}{\mathbf{AG}}{{\mathbf{R}}_{\text{x}}})}\!\!\! \\
							\!\!\!
							{{\operatorname{tr}}( {\mathbf{R}}_{\text{x}}^H{{\mathbf{G}}^H}{{\mathbf{A}}^H}{{\mathbf{D}}_{\text{a}}}{{\mathbf{J}}^H} )}& \!\!\! {{\operatorname{tr}}({{\mathbf{J}}}{\mathbf{AG}}{{\mathbf{R}}_{\text{x}}})}
					\end{array}} \right] \! \succeq \! {\mathbf{0}},\\
		&\eqref{constraint_power}, \eqref{constraint_nonnegative},
		\end{align}
\end{subequations}
which is a standard SDP and is readily to be solved using existing solvers, e.g., CVX. 

\subsubsection{IRS Beamforming Optimization}
For any given $\mathbf{R}_\text{x}$, we define $\mathbf{R}_3 = {\frac{{({K^2} - 1)}}{3}} \mathbf{R}_1 + {{\mathbf{D}}_{\text{a}}}{{\mathbf{R}}_1}{{\mathbf{D}}_{\text{a}}}$. By leveraging the Schur's complement condition with the introduced auxiliary variable $t_2$, problem \eqref{P_new} is transformed into
\begin{subequations}
	\label{P_v}
	\begin{align}
		\mathop {\max }\limits_{{\mathbf{V}},t_2} \;\;\;\; &  t_2\\
		{\text{s.t.}}\;\;\;\;
		&\left[ {\begin{array}{*{20}{c}}
					{\operatorname{tr}\left( \mathbf{R}_3 \mathbf{V}\right) - t_2}&{\operatorname{tr} \left( {{\mathbf{D}_\text{a}}{{\mathbf{R}}_1}{\mathbf{V}}} \right)}\\
					{\operatorname{tr} \left( {{{\mathbf{V}}^H}{\mathbf{R}}_1^H{\mathbf{D}_\text{a}}} \right)}&{\operatorname{tr} \left( {{{\mathbf{R}}_1}{\mathbf{V}}} \right)}
			\end{array}} \right] \succeq {\mathbf{0}}, \\
			&\eqref{constraint_rankV}, \eqref{constraint_rankV},
		\end{align}
\end{subequations}
which is also a standard convex SDP and is solvable with numerical methods available in tools such as CVX. 

\subsubsection{Convergence and Computational Complexity}
The objective value of problem \eqref{P_new} is non-decreasing after each iteration since the sub-problem for $\mathbf{R}_\text{x}$ or $\mathbf{V}$ is solved optimally. Moreover, the optimal objective value has an upper bound from above, ensuring that the proposed algorithm will converge. Upon the AO algorithm reaching convergence, if the solution is not rank-one, the standard Gaussian randomization method can be leveraged to generate candidate solutions that are restricted to be rank-one and feasible to the optimization problem. Then, we choose the feasible solution with the minimum objective value as the final solution and update $\mathbf{R}_\text{x}$. The main computational complexity of the AO algorithm is primarily determined by the solution of the two SDP sub-problems. Specifically, the complexity for solving problems \eqref{TBO} and \eqref{P_v} are $\mathcal{O}\left(M^{3.5}\right)$ and $\mathcal{O}\left(N^{3.5}\right)$, respectively \cite{complexity}. Therefore, the overall complexity of the AO algorithm is given by $\mathcal{O} \left( \left(M^{3.5} + N^{3.5} \right) I_\text{iter} \right)$, where $I_\text{iter}$ denotes the number of iterations needed for convergence.

\section{CRB Analysis and Optimization for Extended Target Case}
\label{Extended Target Case}
In the previous section, the analysis of the CRB of $\theta$ is presented under the point target case. This section is dedicated to exploring the scenarios involving another target model, i.e., the extended target. We first derive the CRB for the TRM estimation, and then the CRB is minimized through the optimization of the BS transmit beamforming.

Under the extended target case, the dedicated active sensors at the IRS have no prior knowledge about the scattering characteristics. Therefore, we estimate the TRM $\mathbf{H}$ instead of $\theta _s$, $\forall s \in \mathcal{S}$.\footnote{The angle and reflection coefficients of each scatter can be extracted from $\mathbf{H}$ using signal processing techniques, e.g., the amplitude and phase estimation of a sinusoid \cite{APES} and generalized likelihood ratio test algorithms \cite{GLRT}.} In this case, the vectorization of ${\mathbf{Y}}$ in \eqref{Y} is denoted by
\begin{align}
{{\mathbf{\tilde y}}} = {\operatorname{vec}} \left( {\mathbf{Y}} \right) = {\mathbf{\tilde z}} + {\mathbf{\tilde n}},
\end{align}
where ${\mathbf{\tilde n}}$, denoting the vectorization of the matrix ${\mathbf{N}}$, is distributed as $\mathcal{CN} ( {{\mathbf{0}},{{\mathbf{\tilde R}}_{\text{n}}}} )$ with ${{\mathbf{\tilde R}}_{\text{n}}}=\sigma _{\text{R}}^2{{\mathbf{I}}_{KT}}$. ${\mathbf{\tilde z}} = \operatorname{vec} \left( {{\mathbf{H\Phi GX}}} \right) = \left( {{{\mathbf{X}}^T}{{\mathbf{G}}^T}{{\mathbf{\Phi }}^T} \otimes {{\mathbf{I}}_K}} \right){\mathbf{h}}$ with $\mathbf{h} = {{\operatorname{vec}}\left( {\mathbf{H}} \right)}$.

\subsection{Estimation Performance Evaluation via CRB}
In this case, we have $2KN$ real parameters to be estimated, which are denoted by ${\bm{\zeta }} = {\left[ {{\mathbf{h}}_{\text{R}}^T,{\mathbf{h}}_{\text{I}}^T} \right]^T}$, where ${{\mathbf{h}}_{\text{R}}} = {\operatorname{Re}} \{\mathbf{h} \}$ and ${{\mathbf{h}}_{\text{I}}} = {\operatorname{Im}} \{ \mathbf{h} \}$. Denote the FIM associated with the estimation of ${\bm{\zeta}}$ by ${\mathbf{\tilde F}} \in {\mathbb{R}^{{2KN} \times {2KN}}}$. Similar to \eqref{F_point}, we can derive each entry of the FIM ${\mathbf{\tilde F}}$. Accordingly, the FIM is expressed as
\begin{align}
\label{FIM_extended}
{\mathbf{\tilde F}} = \left[ {\begin{array}{*{20}{c}}
{{{{\mathbf{\tilde F}}}_{{{\mathbf{h}}_{\text{R}}}{{\mathbf{h}}_{\text{R}}}}}}&{{{{\mathbf{\tilde F}}}_{{{\mathbf{h}}_{\text{R}}}{{\mathbf{h}}_{\text{I}}}}}}\\
{{\mathbf{\tilde F}}_{{{\mathbf{h}}_{\text{I}}}{{\mathbf{h}}_{\text{R}}}}}&{{{{\mathbf{\tilde F}}}_{{{\mathbf{h}}_{\text{I}}}{{\mathbf{h}}_{\text{I}}}}}}
\end{array}} \right],
\end{align}
where
\begin{align}
{{{\mathbf{\tilde F}}}_{{{\mathbf{h}}_{\text{R}}}{{\mathbf{h}}_{\text{R}}}}} &= {{{\mathbf{\tilde F}}}_{{{\mathbf{h}}_{\text{I}}}{{\mathbf{h}}_{\text{I}}}}} = \frac{{2T}}{{\sigma _{\text{R}}^2}}{\operatorname{Re}}\left\{ {\left( {{{\mathbf{\Phi }}^*}{{\mathbf{G}}^*}{{\mathbf{R}}_{\text{X}}^*}{{\mathbf{G}}^T}{{\mathbf{\Phi }}^T}} \right) \otimes {\mathbf{I}}_K} \right\}, \label{F11}\\
{{{\mathbf{\tilde F}}}_{{{\mathbf{h}}_{\text{I}}}{{\mathbf{h}}_{\text{R}}}}} &=  - {{{\mathbf{\tilde F}}}_{{{\mathbf{h}}_{\text{R}}}{{\mathbf{h}}_{\text{I}}}}} \nonumber\\
&= \frac{{2T}}{{\sigma _{\text{R}}^2}}{\operatorname{Im}}\left\{ {\left( {{{\mathbf{\Phi }}^*}{{\mathbf{G}}^*}{{\mathbf{R}}_{\text{X}}^*}{{\mathbf{G}}^T}{{\mathbf{\Phi }}^T}} \right) \otimes {\mathbf{I}}_K} \right\}. \label{F12}
\end{align}
The details of the above derivations can be found in Appendix B. By combining the above analysis, we derive the CRB for the TRM estimation as follows: 

\begin{Proposition}
	\label{propo_ext}
	Under the extended target case, the CRB for the TRM estimation is expressed as
	\begin{align}
		\label{crb_extended}
		{\text{CRB}}( {\mathbf{H}} ) = \frac{{\sigma _{\text{R}}^2K}}{T}{\operatorname{tr}}( {{{( {{\mathbf{G}}{{\mathbf{R}}^H_{\text{x}}}{{\mathbf{G}}^H}} )}^{ - 1}}} ).
	\end{align}
\end{Proposition}

{\it{Proof:}} The CRB of $\mathbf{H}$ can be obtained by taking the inverse of the FIM. Substituting \eqref{F11} and \eqref{F12} into \eqref{FIM_extended}, we can obtain the CRB of $\mathbf{H}$ as
\begin{align}
	{\text{CRB}}\left( {\mathbf{H}} \right) &= \operatorname{tr} \left({{{\mathbf{\tilde F}}}^{ - 1}} \right) \nonumber\\
	&=\frac{{\sigma _{\text{R}}^2}}{T}{\operatorname{tr}}\left( {{{\left( {{\mathbf{\Phi G}}{{\mathbf{R}}^H_{\text{x}}}{{\mathbf{G}}^H}{{\mathbf{\Phi }}^H}} \right)}^{ - 1}}} \right){\operatorname{tr}}\left( {{\mathbf{I}}_K^{ - 1}} \right) \nonumber\\
	&= \frac{{\sigma _{\text{R}}^2K}}{T}{\operatorname{tr}}\left( {{{\mathbf{\Phi }}^H}{{\left( {{\mathbf{G}}{{\mathbf{R}}^H_{\text{x}}}{{\mathbf{G}}^H}} \right)}^{ - 1}}{\mathbf{\Phi }}} \right) \nonumber\\
	&= \frac{{\sigma _{\text{R}}^2K}}{T}{\operatorname{tr}}\left( {{{\left( {{\mathbf{G}}{{\mathbf{R}}^H_{\text{x}}}{{\mathbf{G}}^H}} \right)}^{ - 1}}} \right),
\end{align}
which completes the proof. ~$\hfill\blacksquare$

Proposition \ref{propo_ext} shows that the CRB for the TRM estimation does not rely on the design of the IRS reflective beamforming. By comparing \eqref{CRB_point} and \eqref{crb_extended} presented in Proposition \ref{Pro_point} and Proposition \ref{propo_ext}, one can observe that the CRB increases with $\sigma_\text{R}^2$ and decreases with $T$ under both cases. Moreover, it can be seen that more sensors degrade the sensing performance under the extended target case, whereas it results in a lower CRB under the point target case, which is explained in Section \ref{Point Target Case}. This is because increasing $K$ leads to more parameters that need to be estimated, which means more unknowns to be determined from the available information. The available information is dispersed among more unknowns, increasing estimation uncertainty, which makes it more challenging to accurately estimate the parameters.

According to \cite{FPS}, the CRB for the TRM estimation of the fully-passive IRS-enabled sensing system is given by ${\text{CRB}_{{\text{FPS}}}}\left( {\mathbf{H}} \right) = \frac{{\sigma _{\text{R}}^2}}{T}{\operatorname{tr}}\left( {{{\left( {{\mathbf{G}}{{\mathbf{R}}^H_{\text{x}}}{{\mathbf{G}}^H}} \right)}^{ - 1}}} \right){\operatorname{tr}}\left( {{{\left( {{\mathbf{G}_\text{r}^H}{{\mathbf{G}}_\text{r}}} \right)}^{ - 1}}} \right)$, where $\mathbf{G}_\text{r} \in \mathbb{C}^{M_\text{r} \times N}$ denotes the IRS-BS channel with the number of receive antenna $M_\text{r}$. Comparing the CRB for the TRM estimation of the two sensing systems with different IRS architectures, we have the following remarks.
\begin{Remark}
It is worth mentioning that the FIM ${\mathbf{\tilde F}}$ in \eqref{FIM_extended} is a singular matrix, and the corresponding CRB is infinite when $\operatorname{rank}\left(\mathbf{G}\right) \textless N$. This is because the lack of sufficient degrees of freedom to accurately measure the target's direction limits the precision of the DoA estimation. As such, in both fully-passive and semi-passive IRS-enabled sensing systems, the common requirement for $\mathbf{H}$ to be estimable under the extended target case is that $\operatorname{rank}\left(\mathbf{G}\right) = N$. However, there exists an extra condition to be fulfilled in the fully-passive IRS-enabled sensing system, i.e., $\operatorname{rank}\left(\mathbf{G}_\text{r}\right) \textless N$ \cite{FPS}.
\end{Remark}
\begin{Remark}
\label{remark_extended}
Under the extended target case, the CRB of the sensing system with the semi-passive IRS outperforms that with the fully-passive IRS if $K \textless {\operatorname{tr}}\left( {{{\left( {{\mathbf{G}_\text{r}^H}{{\mathbf{G}}_\text{r}}} \right)}^{ - 1}}} \right)$. This is because a lower CRB implies that the former architecture offers better estimation accuracy, as it has a smaller theoretical lower bound on the variance. By applying SVD to $\mathbf{G}_\text{r}$, the expression ${\operatorname{tr}}\left( {{{\left( {{\mathbf{G}_\text{r}^H}{{\mathbf{G}}_\text{r}}} \right)}^{ - 1}}} \right)$ can be rewritten as the sum of the reciprocals of the squared singular values of $\mathbf{G}_\text{r}$, which is generally much larger than $K$.
\end{Remark}

\subsection{Transmit Beamforming Design}
Given the derived CRB for the TRM estimation in \eqref{crb_extended}, there is no need to optimize the reflective beamformer $\mathbf{\Phi }$. Thus, the CRB minimization problem by optimizing ${\mathbf{R}}_{\text{x}}$ is formulated as
\begin{subequations}
\label{extended}
\begin{align}
\mathop {\min }\limits_{{{\mathbf{R}}_{\text{x}}}} \;\;\;\; &{\operatorname{tr}}\left( {{{\left( {{\mathbf{G}}{{\mathbf{R}}^H_{\text{x}}}{{\mathbf{G}}^H}} \right)}^{ - 1}}} \right) \label{obj_ext}\\
{\text{s.t.}}\;\;\;\;
&{\operatorname{tr}} \left( {{{\mathbf{R}}_{\text{x}}}} \right) \le {P_0}, \label{constraint_P0_ext}\\
&{{\mathbf{R}}_{\text{x}}} \ge {\mathbf{0}} \label{constraint_neg_ext}.
\end{align}
\end{subequations}
Note that the objective function \eqref{obj_ext} and constraints \eqref{constraint_P0_ext} and \eqref{constraint_neg_ext} are convex, thus making problem \eqref{extended} convex. Then, we derive the optimal solution and the CRB in closed-form as follows. We denote the SVD of ${\mathbf{G}}$ as ${\mathbf{G}} = {\mathbf{\hat U} \mathbf{\hat \Sigma} \mathbf{\hat Q}^{H}}$, where ${\mathbf{\hat U}} \in \mathbb{R}^{N \times N}$ and ${\mathbf{\hat Q}} \in \mathbb{R}^{M \times M}$ with ${\mathbf{\hat U}^H}{{{\mathbf{\hat U}}}} = {\mathbf{\hat U}}{{{\mathbf{\hat U}}}^H} = {{\mathbf{I}}_N}$, ${\mathbf{\hat Q}^H}{{{\mathbf{\hat Q}}}} = {\mathbf{\hat Q}}{{{\mathbf{\hat Q}}}^H} = {{\mathbf{I}}_M}$, and ${\mathbf{\hat \Sigma }} = \operatorname{diag} \left( {{{\hat \sigma }_1}, \ldots ,{{\hat \sigma }_N}}, 0, \ldots,0 \right) = [ {{{{\mathbf{\hat \Sigma }}}_1},{\mathbf{0}}} ] \in {\mathbb{R}^{N \times M}}$ with ${{\hat \sigma }_1} \ge  \ldots  \ge {{\hat \sigma }_N} \textgreater 0$ denoting the $N$ positive singular values. Following a similar procedure as in \cite{FPS}, the optimal solution to problem \eqref{extended} is given by 
\begin{align}
	\label{Rx_extended}
	{\mathbf{R}}_{\text{x}}^* = {\mathbf{\hat Q\hat R}}_{\text{x}}^{{\text{opt}}}{{{\mathbf{\hat Q}}}^H},	
\end{align}
where
\begin{align}
	{\mathbf{\hat R}}_{\text{x}}^{{\text{opt}}} = \left[ {\begin{array}{*{20}{c}}
			{\frac{{{\mathbf{\hat \Sigma }}_1^{ - 1}{P_0}}}{{\sum\nolimits_{i = 1}^N {\hat \sigma _i^{ - 1}} }}}&{\mathbf{0}}\\
			{\mathbf{0}}&{\mathbf{0}}
	\end{array}} \right].
\end{align}
Based on \eqref{crb_extended} and \eqref{Rx_extended}, the optimal CRB for the TRM estimation under the extended target case is expressed as
\begin{align}
	\label{crb_opt_ext}
	{\text{CRB}}{\left( {\mathbf{H}} \right)^{{\text{opt}}}} = \frac{{\sigma _{\text{R}}^2K}}{{{P_0}T}}{{\left( {\sum\nolimits_{i = 1}^N {\hat \sigma _i^{ - 1}} } \right)^2}}.
\end{align}
From \eqref{crb_opt_ext}, it is noticed that the value of the inverse singular value $\hat \sigma _i^{ - 1}$ depends on both $M$ and $N$, and the number of $\hat \sigma _i^{ - 1}$ increases with $N$. The impact of $M$ and $N$ on the CRB will be explored in detail in Section \ref{Numerical results}.

\section{Simulation results}
\label{Numerical results}
To characterize the performance of our proposed CRB-based scheme in the semi-passive IRS-enabled sensing system, numerical results are presented in this section. The distance between BS and IRS is set as $d_\text{BI} = 60$ meter (m), and that between the IRS and target is $d_\text{IT} = 20$ m. Denote the large-scale path loss model by $L\left(d\right) = C_0\left({d}/{D_0}\right)^{-\alpha _d}$ with the path-loss exponent $\alpha_d$, the link distance $d$, and the path loss $C_0 = 30$ dB at the reference distance $D_0 = 1$ m. Considering the small-scale fading, we characterize the BS to IRS channel by Rician fading, which is modeled as
\begin{align}	
	\mathbf{G} = \rho_\text{BI} \left( \sqrt{\frac{\beta_\text{BI}}{\beta_\text{BI}+1}} {\mathbf{\bar G}} + \sqrt{\frac{1}{\beta_\text{BI}+1}} {\mathbf{\tilde G}} \right),
\end{align}
where $\rho_\text{BI}$ is the large-scale path-loss with the path-loss exponent $\alpha_\text{BI}$, and $\beta_\text{BI}$ denotes the Rician factor. $\mathbf{\bar G}$ and $\mathbf{\tilde G}$ denote the LoS and NLoS components of the channel, respectively. We set $\beta_\text{BI} = 5$ dB and $\alpha_\text{BI} = 2.5$. Unless otherwise specified, other system parameters given in the following are set as $\theta = 60^\circ$, $\kappa = 7$ dBsm, $\lambda_{\text{R}} = 0.2$ m, $\hat{d} = \lambda_{\text{R}}/2$, $\sigma_{\text{R}}^2 = -90$ dBm, and $T = 64$. 

\subsection{Point Target Case}
\begin{figure}[t]
	\centering
	\includegraphics[width=0.48\textwidth]{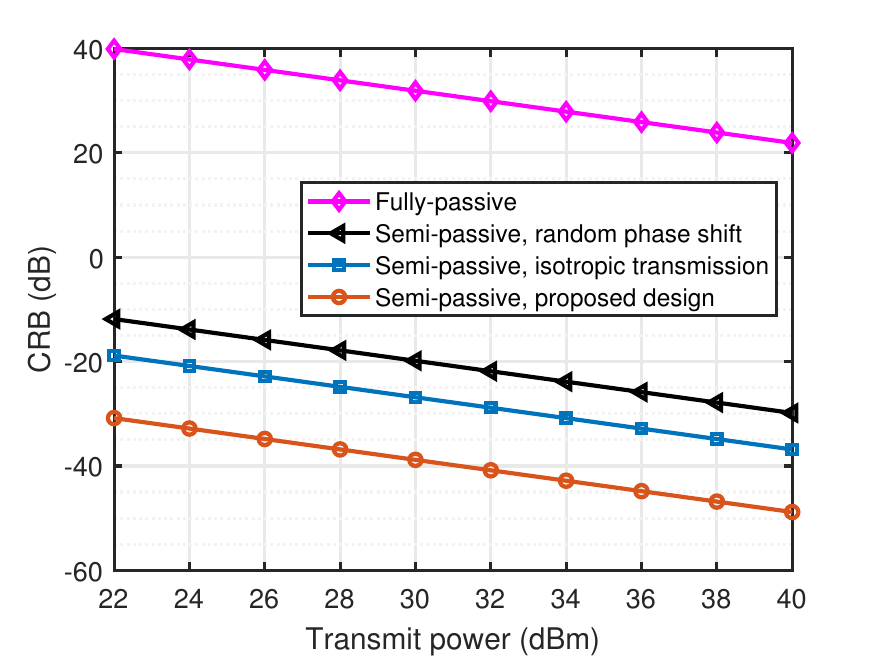}
	\caption{The CRB for the DoA estimation versus $P_0$ for different schemes.}
	\label{fig:Pt_RMSE}
	\vspace{-5pt}
\end{figure}
To illustrate the effectiveness of the proposed scheme with the semi-passive IRS, the following schemes are considered for comparison: \textbf{1) Fully-passive \cite{FPS}:} The number of receive antennas at the BS $M_\text{r}$ is set equal to the number of sensors $K$ for a fair comparison by assuming that no active sensors are deployed at the IRS; \textbf{2) Semi-passive, random phase shift:} the IRS phase shifts are randomly chosen over the range $\left[0, 2\pi \right)$, and the CRB is minimized by optimizing the BS beamforming; \textbf{3) Semi-passive, isotropic transmission:} the BS transmits orthonormal signal beams, i.e., $\mathbf{R}^\text{iso}_\text{x} = P_0 \mathbf{I}_\text{M}/M$ \cite{iso}, while the phase shifts are optimized at the semi-passive IRS; \textbf{4) Semi-passive, proposed design:} the CRB is obtained by our proposed joint beamforming design. In Fig. \ref{fig:Pt_RMSE}, we plot the CRB for the DoA estimation versus $P_0$ when $M_\text{r}=M=K=N=16$. One can observe that the semi-passive IRS architecture outperforms the fully-passive IRS architecture. This can be attributed to the shorter signal propagation path in the former configuration, in which the signal traverses the BITS link. In contrast, the fully-passive IRS-enabled sensing system introduces significant signal attenuation through the BITIB link, which degrades the sensing performance. In addition, for the semi-passive IRS-enabled sensing system, it is observed that the scheme with random phase shifts exhibits inferior performance compared to the schemes with reflective beamforming design (i.e., isotropic transmission and proposed design). This discrepancy can be attributed to the absence of passive beamforming gain, which highlights the importance of carefully designing the IRS beamforming. For the entire range of transmit power, our proposed scheme yields a substantial enhancement in performance compared to other baseline schemes, which demonstrates its superiority in terms of the CRB.

\subsubsection{Impact of the Rician Factor}
\begin{figure}[t]
	\centering
	\includegraphics[width=0.48\textwidth]{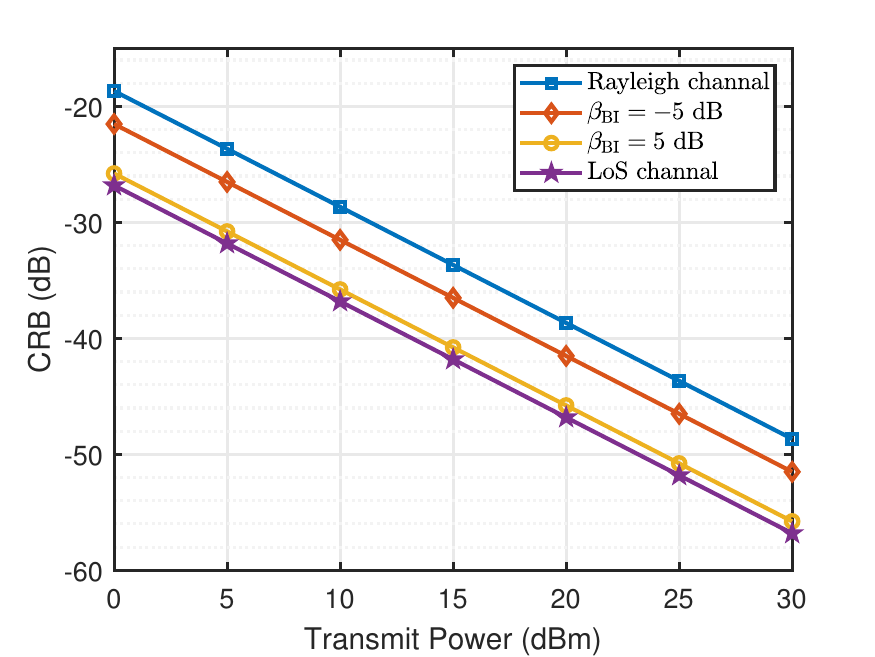}
	\caption{The CRB for the DoA estimation versus $P_0$ under different channel models.}
	\label{fig:Rician}
	\vspace{-5pt}
\end{figure}
In Fig. \ref{fig:Rician}, we investigate the impact of the Rician factor by plotting the CRB for the DoA estimation versus the transmit power $P_0$ under different channel models when $N = 64$, $M = 8$, and $K = 8$. One can observe that the CRB of $\theta$ decreases as $\beta_\text{BI}$ increases. Note that the NLoS component of the BS-IRS channel reduces the beamforming gain, thereby degrading the sensing performance. Semi-passive IRS can be utilized to mitigate this adverse effect and enhance the sensing performance in challenging scenarios. The results highlight the importance of deploying a semi-passive IRS to establish the dominant LoS path, especially when the sensing target is situated in the NLoS area of the BS.

\subsubsection{Impact of the System Parameters}
\begin{figure}[t]
	\centering
	\includegraphics[width=0.48\textwidth]{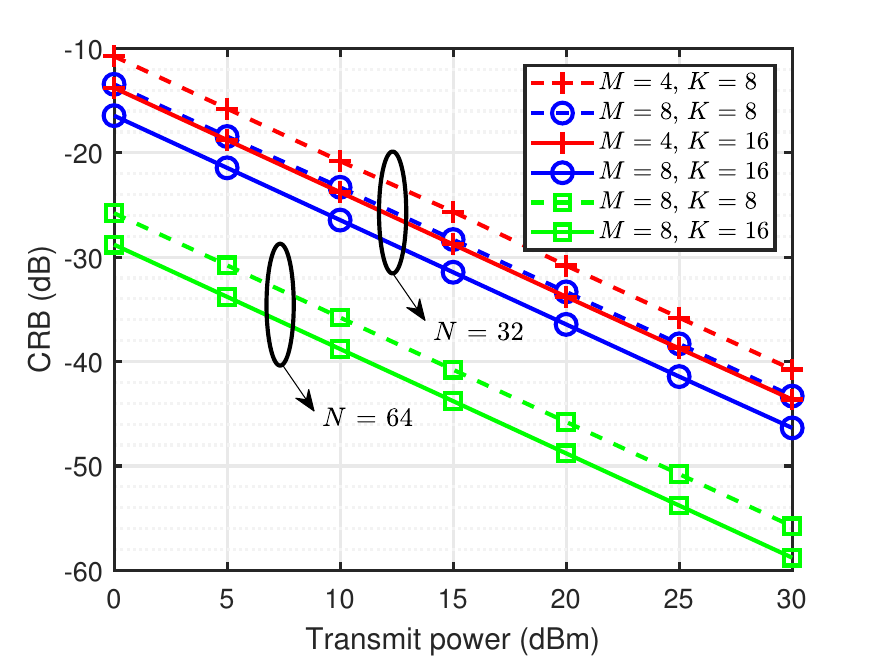}
	\caption{Impact of the system parameters on the CRB for the DoA estimation.}
	\label{fig:KMN}
	\vspace{-5pt}
\end{figure}
In Fig. \ref{fig:KMN}, we evaluate the performance of our proposed design by plotting the $\text{CRB}$ versus $P _0$ under different numbers of antennas $M$, IRS reflecting elements $N$, and sensors $K$. First, one can see that the CRB of $\theta$ decreases linearly with $P _0$ because the CRB is inversely proportional to $P _0$ in \eqref{crb}. Moreover, increasing $P _0$ allows a stronger signal to be transmitted. Second, we can observe that the CRB decreases with $M$ since adding antennas at the BS can provide more spatial diversity and higher beam gain. Third, it can be seen that the CRB decreases with $N$ because an increase in $N$ results in higher passive beamforming gain, thereby contributing to better sensing performance. Finally, we can observe that the CRB decreases with $K$ due to higher spatial resolution and higher array gain, which is consistent with Proposition \ref{CRB_K}. Given $M$, the sensing performance can be effectively enhanced by deploying more sensors at the semi-passive IRS in a flexible manner, except in the fully-passive IRS-enabled sensing system where it can only be achieved by adjusting $N$.

\subsubsection{Impact of the Weight on IRS Elements and Sensors Allocation}
\begin{figure*}[t]
	\vspace{-5pt}
	\centering
	\subfloat[The proposed optimized $\tilde{N}$ and $\tilde{K}$ versus $W_\text{I}$ under different $Q_\text{tot}$.]{
		\centering
		\label{fig:NK_a}
		\includegraphics[width=0.48\textwidth]{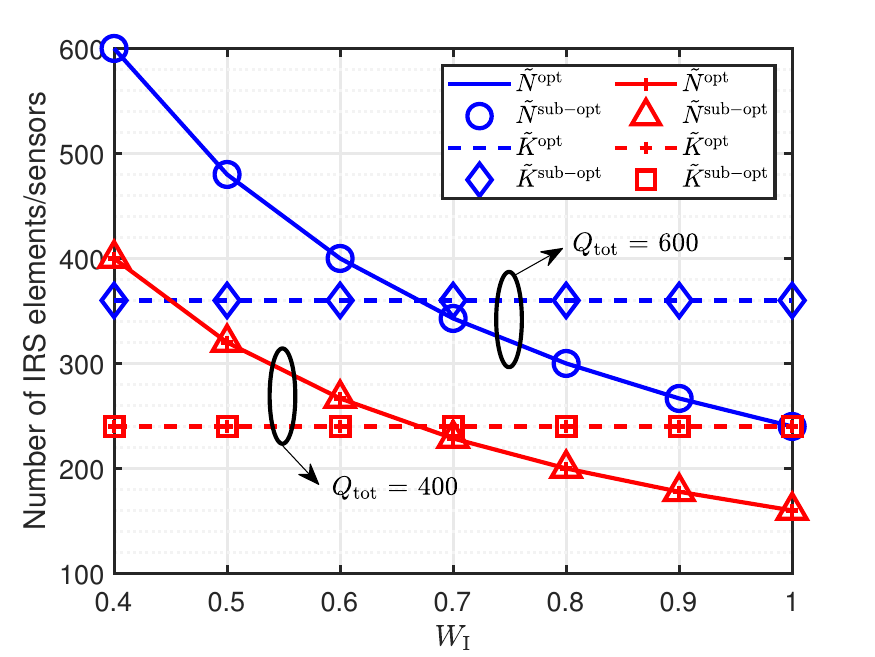}}
	\subfloat[Comparison of the proposed optimized $\tilde{N}$ and $\tilde{K}$ and the solution obtained by the exhaustive search algorithm versus $W_\text{I}$ with $Q_\text{tot} = 600$.]{
		\centering
		\label{fig:NK_b}
		\includegraphics[width=0.48\textwidth]{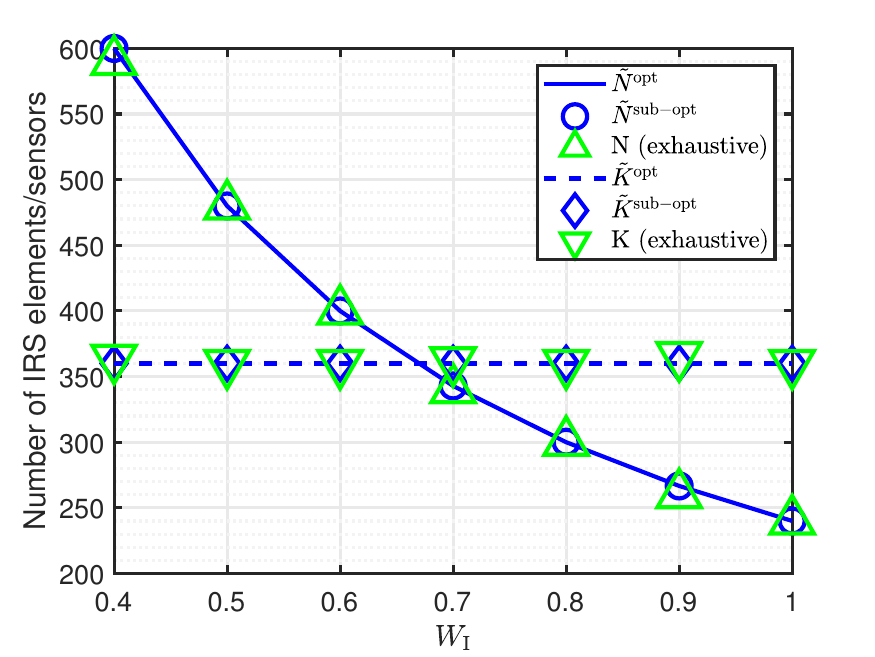}}
	\caption{The optimized number of IRS passive reflecting elements/active sensors versus $W_\text{I}$ with $W_\text{s}=1$.}
	\label{fig:NK}
	\vspace{-5pt}
\end{figure*}
In Fig. \ref{fig:NK_a}, we plot the optimized IRS elements and sensors allocation design for CRB minimization versus $W_\text{I}$ under the special case with a single-antenna BS where $W_\text{s}=1$. One can observe that more reflecting elements should be deployed at the IRS when $W_\text{I} \textless 0.7$. In the regime where $W_\text{I}$ is small, the performance of sensors is mainly restricted by the deployment cost. Compared to the sensors, the reflecting elements enjoy the advantages of lower cost and reaping a higher passive beamforming gain of $\mathcal{O}(N^2)$. Nevertheless, the number of reflecting elements gradually decreases as $W_\text{I}$ increases since the sensors can provide a gain of $\mathcal{O}(K^3)$ including both the spatial direction gain and the receive beamforming gain. As $W_\text{I}$ becomes large, the benefit from the gain provided by the sensors becomes dominant. Moreover, we observe that $\tilde{K}$ is always greater than $\tilde{N}$ when given the total number, i.e., $W_\text{I}=W_\text{s}=1$ for both $Q_\text{tot} = 600$ and $Q_\text{tot} = 400$, which corroborates our analysis in Section \ref{Point Target Case}. Fig. \ref{fig:NK_b} compares the optimized numbers of IRS elements/sensors under different $W_\text{I}$ with $Q_\text{tot} = 600$. It is observed that both the optimal and sub-optimal solutions we derived closely approximate the optimal solution achieved through the exhaustive search algorithm, exhibiting a similar trend. The results illustrate the effectiveness of our proposed solutions.

\subsection{Extended Target Case}
To illustrate the effectiveness of the semi-passive IRS-enabled sensing system with the proposed scheme, we compare the transmit beamforming design with ${\mathbf{R}}_{\text{x}}^*$ in \eqref{Rx_extended} and that with the isotropic transmission solution, in terms of the CRB. With $\mathbf{R}^\text{iso}_\text{x} = P_0 \mathbf{I}_\text{M}/M$ \cite{iso}, the CRB for the TRM estimation of the isotropic transmission scheme can be expressed as
\begin{align}
	\label{crb_iso_ext}
	{\text{CRB}}{\left( {\mathbf{H}} \right)^{{\text{iso}}}} = \frac{{\sigma _{\text{R}}^2KM}}{{{P_0}T}} {{\sum\nolimits_{i = 1}^N {\hat \sigma _i^{ - 2}} }}.
\end{align}
Based on \eqref{crb_opt_ext} and \eqref{crb_iso_ext}, the performance gap in decibels (dB) between the two schemes is given by
\begin{align}
	\label{gap_dB}
	&10\operatorname{lg}({\text{CRB}}{\left( {\mathbf{H}} \right)^{{\text{iso}}}}) - 10\operatorname{lg}({\text{CRB}}{\left( {\mathbf{H}} \right)^{{\text{opt}}}}) \nonumber\\
	=& 10\operatorname{lg}\left({{M\sum\nolimits_{i = 1}^N {\hat \sigma _i^{ - 2}} }}/{{{{\left( {\sum\nolimits_{i = 1}^N {\hat \sigma _i^{ - 1}} }\right)}^2}}}\right).
\end{align}

\subsubsection{Impact of the Transmit Power}
\begin{figure}[t]
	\vspace{-5pt}
	\centering
	\includegraphics[width=0.48\textwidth]{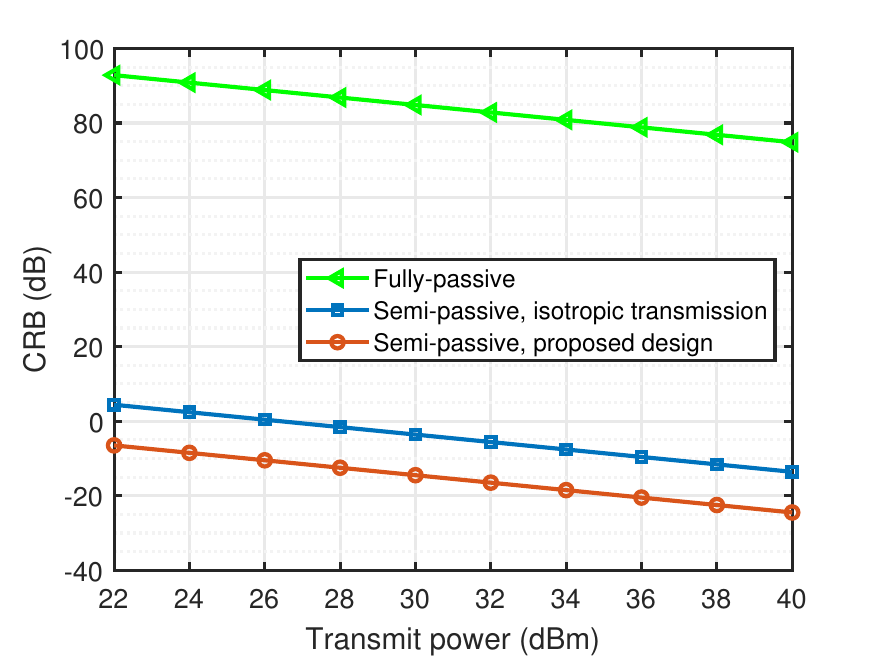}
	\caption{The CRB for the TRM estimation versus $P_0$.}
	\label{fig:Power_CRB_ext}
	\vspace{-10pt}
\end{figure}
Fig. \ref{fig:Power_CRB_ext} shows the CRB comparison versus different transmit power $P_0$ under the extended target case. For a fair comparison, we set $M_\text{r}=M=K=N=32$. One can observe that the CRB decreases linearly with $P_0$ due to its inverse proportionality to $P_0$ in both \eqref{crb_opt_ext} and \eqref{crb_iso_ext}, exhibiting a similar impact as observed in the scenario with a point target. In addition, it can be seen that the performance gap between our proposed CRB-based scheme and the isotropic transmission scheme is constant with the transmit power. This is expected because the performance gap in \eqref{gap_dB} is regardless of $P_0$. For the entire range of transmit power, the results indicate significant performance enhancements with the deployment of the semi-passive IRS, measured by the CRB. Moreover, our proposed scheme consistently demonstrates superior performance, which highlight its superiority.

\subsubsection{Impact of the Number of Sensors at the IRS}
\begin{figure*}[t]
	\vspace{-5pt}
	\centering
	\subfloat[The CRB versus $K$ for the proposed design.]{
		\centering
		\label{fig:K_a}
		\includegraphics[width=0.48\textwidth]{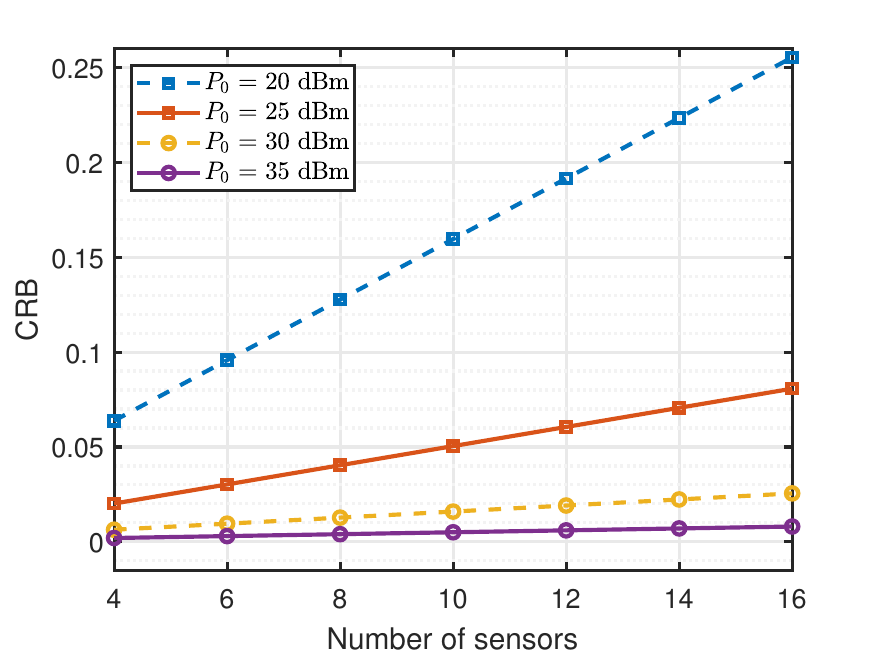}}	
	\subfloat[The CRB versus $K$ for isotropic transmission scheme and proposed design.]{
		\centering
		\label{fig:K_b}
		\includegraphics[width=0.48\textwidth]{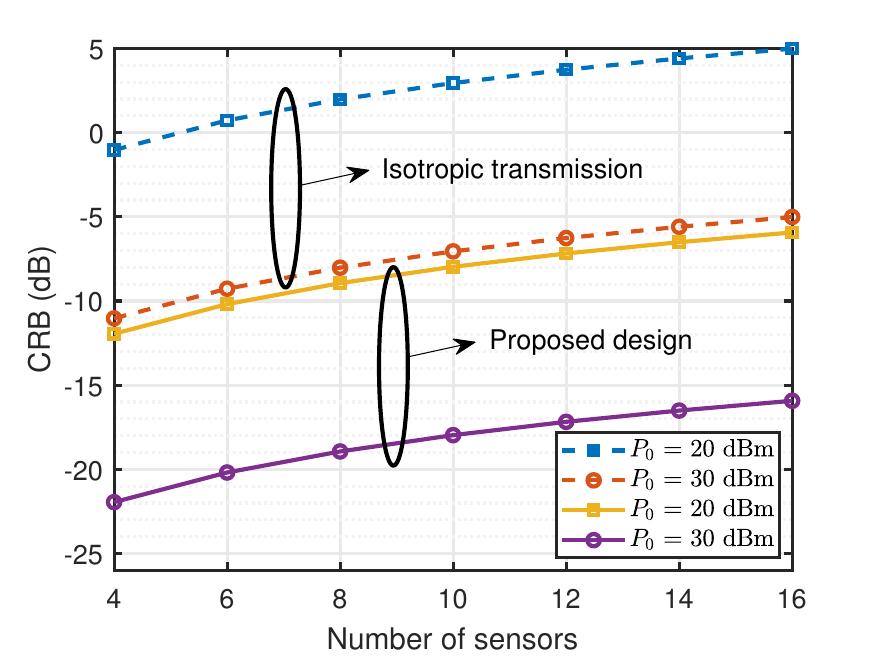}}
	\caption{The CRB for the TRM estimation versus $K$.}
	\label{fig:K}
	\vspace{-10pt}
\end{figure*}
In Fig. \ref{fig:K_a}, we plot the CRB versus $K$ with $M=N=32$ under different transmit power $P_0$. One can observe that the CRB increases linearly with $K$ as revealed in Proposition \ref{propo_ext}. This is because more parameters need to be estimated. Consequently, it leads to an increase in estimation uncertainty, thereby making it more difficult to accurately estimate each parameter. Nevertheless, it is observed that the adverse effects caused by an increase in $K$ on CRB can be mitigated by increasing $P_0$. This is due to the fact that increasing $P_0$ is helpful for augmenting the power received at the sensors. In Fig. \ref{fig:K_b}, we compare the proposed scheme with the isotropic transmission scheme by plotting the CRB in dB versus $K$ with $M=N=32$ for both $P_0 = 20$ dBm and $P_0 = 30$ dBm. Finally, we observe that our proposed design performs better, and the performance gap in dB between the two schemes remains constant, which agrees with the analysis of \eqref{gap_dB}, i.e., $10\operatorname{lg}({\text{CRB}}{( {\mathbf{H}} )^{{\text{iso}}}}) - 10\operatorname{lg}({\text{CRB}}{( {\mathbf{H}} )^{{\text{opt}}}})$ is regardless of $K$.

\subsubsection{Impact of the Number of Antennas at the BS}
\begin{figure}[t]
	\centering
	\includegraphics[width=0.48\textwidth]{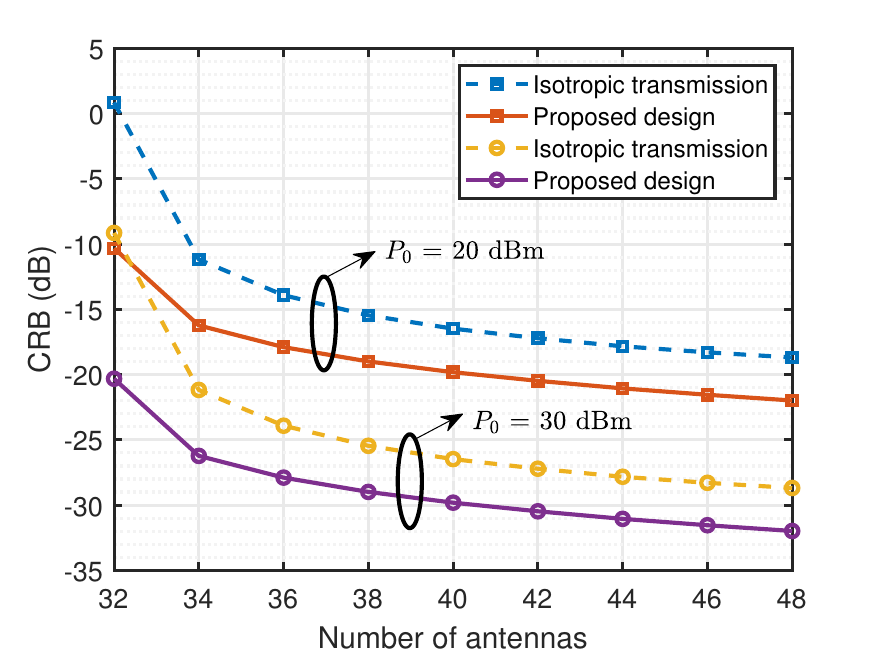}
	\caption{The CRB for the TRM estimation versus $M$.}
	\label{fig:M_ext}
	\vspace{-10pt}
\end{figure}
In Fig. \ref{fig:M_ext}, we plot the CRB versus $M$ when $N=32$ and $K=8$. One can see that the CRB decreases with $M$ for both $P_0 = 20$ dBm and $P_0 = 30$ dBm. The reason is that the increase in $M$ leads to an increase in the sum of singular values and thus a decrease in the sum of inverse singular values, i.e., $( {\sum\nolimits_{i = 1}^N {\hat \sigma _i^{ - 1}} } )^2$. Intuitively, more transmit antennas deployed at the BS can provide more spatial diversity and higher beam gain, which is explained under the point target case. In addition, we observe that the proposed design achieves a lower CRB, and the performance gap between them first decreases and then increases with $M$. For example, for $P_0 = 30$ dBm, the performance gap is 11.18 dB, 3.26 dB, and 3.3 dB when $M$ is 32, 42, and 48, respectively. For fixed $N$, ${{\sum\nolimits_{i = 1}^N {\hat \sigma _i^{ - 2}} }}/{{{{( {\sum\nolimits_{i = 1}^N {\hat \sigma _i^{ - 1}} } )}^2}}}$ roughly decreases with $M$ because the distribution of singular values tends to become more spread out and diverse. From \eqref{gap_dB}, the performance gap decreases with $M$ when $M$ is not very large and ${{\sum\nolimits_{i = 1}^N {\hat \sigma _i^{ - 2}} }}/{{{{( {\sum\nolimits_{i = 1}^N {\hat \sigma _i^{ - 1}} } )}^2}}}$ dominates. Moreover, the CRB gap becomes larger with $M$ because $M$ dominates.

\subsubsection{Impact of the Number of IRS Reflecting Elements}
\begin{figure}[t]
	\centering
	\includegraphics[width=0.48\textwidth]{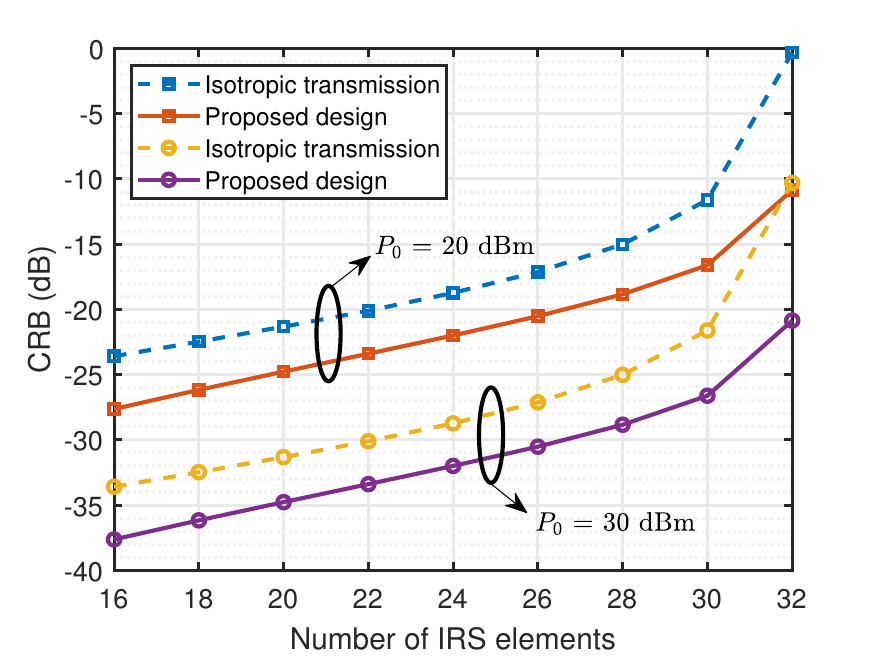}
	\caption{The CRB for the TRM estimation versus $N$.}
	\label{fig:N_ext}
	\vspace{-5pt}
\end{figure} 
In Fig. \ref{fig:N_ext}, we investigate the impact of $N$ based on our proposed design by plotting the CRB versus $N$ when $M = 32$ and $K = 8$. One can observe that the CRB increases with $N$ for both $P_0 = 20$ dBm and $P_0 = 30$ dBm. This is because ${\sum\nolimits_{i = 1}^N {\hat \sigma _i^{ - 1}} }$ increases with $N$ in \eqref{crb_opt_ext}, which leads to a higher CRB. The results differ from those under the point target case because we analyze the CRB of $\mathbf{H}$ under the extended target case. In this case, increasing $N$ yields more parameters that need to be estimated, which is similar to the argument for the influence of the number of IRS sensors. In addition, the CRB achieved by the proposed design is lower than the isotropic transmission scheme, and the gap between them first decreases and then increases with $N$. For example, for $P_0 = 30$ dBm, the performance gap is 4.04 dB, 3.4 dB, and 10.56 dB when $N$ is 16, 26, and 32, respectively. For fixed $M$, both $\sum\nolimits_{i = 1}^N {\hat \sigma _i^{ - 2}}$ and ${( {\sum\nolimits_{i = 1}^N {\hat \sigma _i^{ - 1}} } )^2}$ in \eqref{gap_dB} increase with $N$. The square of the sum, i.e., ${( {\sum\nolimits_{i = 1}^N {\hat \sigma _i^{ - 1}} } )^2}$, increases significantly and dominates when $N$ is not very large, resulting in the performance gap that decreases with $N$. $M \sum\nolimits_{i = 1}^N {\hat \sigma _i^{ - 2}}$ dominates when $N$ becomes larger and close to $M$, thus the performance gap is more pronounced.

\section{Conclusion}
\label{Conclusion}
In this paper, we investigated the CRB minimization problem in a semi-passive IRS-enabled sensing system. In particular, we first derived the CRB and formulated the corresponding minimization problems for both the point and extended targets. Under the point target case with a single-antenna BS, the IRS reflecting elements and sensors allocation design based on the optimal CRB for the DoA estimation was characterized. For the case with a multi-antenna BS, an efficient algorithm was developed for tackling the optimization problem. Under the extended target case, we minimized the CRB for the TRM estimation through the optimization of the BS transmit beamforming. Numerical results validated that, compared with the fully-passive IRS, the semi-passive IRS can significantly enhance the sensing performance. Moreover, the impact of the system parameters on the CRB was investigated. Specifically, more reflecting elements and sensors should be installed at the IRS when dealing with the point target, whereas fewer are preferable for scenarios involving an extended target. Besides, increasing the number of antennas and the transmit power is beneficial for sensing in both cases. The results in this paper showed the superiority of our proposed CRB-based scheme and illustrated that integrating it into semi-passive IRS-enabled sensing systems is a promising solution for target estimation.

\section*{Appendix A}
\label{AppendixA}
Since the noise is Gaussian distributed, the covariance matrix ${{{\mathbf{\hat R}}_{\text{n}}}}$ is independent of ${\bm{\varsigma }}$. According to \eqref{F_point}, we have
\begin{align}
\frac{{\partial {\mathbf{\hat z}}}}{{\partial \theta }} &= \alpha \operatorname{vec} \left( {{\mathbf{\dot EX}}} \right),\\
\frac{{\partial {\mathbf{\hat z}}}}{{\partial {\bm{\hat \alpha }}}} &= \left[ {1,j} \right] \otimes \operatorname{vec} \left( {{\mathbf{EX}}} \right).
\end{align}
Then, ${{{\hat F}}_{\theta \theta }}$, ${{\mathbf{\hat F}}_{\theta {\bm{\hat \alpha }}}}$ and ${{\mathbf{\hat F}}_{{\bm{\hat \alpha \hat \alpha }}}}$ in \eqref{FIM_point} are given by
\begin{align}
	{{{\hat F}}_{\theta \theta }} &= \frac{2}{{\sigma _{\text{R}}^2}}{\operatorname {Re}} \left\{ {{{\left( {\alpha \operatorname{vec} \left( {{\mathbf{\dot EX}}} \right)} \right)}^H} \cdot \left( {\alpha \operatorname{vec} \left( {{\mathbf{\dot EX}}} \right)} \right)} \right\} \nonumber\\
	&= \frac{{2T{{\left| \alpha  \right|}^2}}}{{\sigma _{\text{R}}^2}}\operatorname{tr} \left( {{\mathbf{\dot E}}{{\mathbf{R}}_{\text{x}}}{{{\mathbf{\dot E}}}^H}} \right),
\end{align}
\begin{align}
{{\mathbf{\hat F}}_{\theta {\bm{\hat \alpha }}}} &= \frac{2}{{\sigma _{\text{R}}^2}}{\operatorname {Re}} \left\{ {{{\left( {\alpha \operatorname{vec} \left( {{\mathbf{\dot EX}}} \right)} \right)}^H} \cdot \left( {\left[ {1,j} \right] \otimes \operatorname{vec} \left( {{\mathbf{EX}}} \right)} \right)} \right\} \nonumber\\
&= \frac{{2T}}{{\sigma _{\text{R}}^2}}{\operatorname {Re}} \left\{ {{\alpha ^*}\operatorname{tr} \left( {{\mathbf{E}}{{\mathbf{R}}_{\text{x}}}{{{\mathbf{\dot E}}}^H}} \right)\left[ {1,j} \right]} \right\},
\end{align}
\begin{align}
{{\mathbf{\hat F}}_{{\bm{\hat \alpha \hat \alpha }}}} &= \frac{2}{{\sigma _{\text{R}}^2}}{\operatorname {Re}} \left\{ {{{\left( {\left[ {1,j} \right] \otimes \operatorname{vec} \left( {{\mathbf{EX}}} \right)} \right)}^H} \cdot \left( {\left[ {1,j} \right] \otimes \operatorname{vec} \left( {{\mathbf{EX}}} \right)} \right)} \right\} \nonumber\\ 
&= \frac{{2T}}{{\sigma _{\text{R}}^2}}\operatorname{tr} \left( {{\mathbf{E}}{{\mathbf{R}}_{\text{x}}}{{\mathbf{E}}^H}} \right){{\mathbf{I}}_2}.	
\end{align}

\section*{Appendix B}
\label{AppendixB}
Since the noise is Gaussian distributed, the covariance matrix ${{{\mathbf{\tilde R}}_{\text{n}}}}$ is independent of ${\bm{\zeta }}$. Consequently, we have the following partial derivatives:
\begin{align}
&\frac{{\partial {\mathbf{\tilde z}}}}{{\partial {{\mathbf{h}}_{\text{R}}}}} = {{\mathbf{X}}^T}{{\mathbf{G}}^T}{{\mathbf{\Phi }}^T} \otimes {{\mathbf{I}}_K},\\
&\frac{{\partial {\mathbf{\tilde z}}}}{{\partial {{\mathbf{h}}_{\text{I}}}}} = j{{\mathbf{X}}^T}{{\mathbf{G}}^T}{{\mathbf{\Phi }}^T} \otimes {{\mathbf{I}}_K}.
\end{align}
Then, ${{{\mathbf{\tilde F}}}_{{{\mathbf{h}}_{\text{R}}}{{\mathbf{h}}_{\text{R}}}}}$, ${{{\mathbf{\tilde F}}}_{{{\mathbf{h}}_{\text{I}}}{{\mathbf{h}}_{\text{I}}}}}$, ${{{\mathbf{\tilde F}}}_{{{\mathbf{h}}_{\text{I}}}{{\mathbf{h}}_{\text{R}}}}}$ and ${{{\mathbf{\tilde F}}}_{{{\mathbf{h}}_{\text{R}}}{{\mathbf{h}}_{\text{I}}}}}$ in \eqref{FIM_extended} are given by
\begin{align}
{{{\mathbf{\tilde F}}}_{{{\mathbf{h}}_{\text{R}}}{{\mathbf{h}}_{\text{R}}}}} &= {{{\mathbf{\tilde F}}}_{{{\mathbf{h}}_{\text{I}}}{{\mathbf{h}}_{\text{I}}}}} \nonumber\\
&= \frac{{2T}}{{\sigma _{\text{R}}^2}}{\text{Re}}\left\{ {{{( {{{\mathbf{X}}^T}{{\mathbf{G}}^T}{{\mathbf{\Phi }}^T} \otimes {{\mathbf{I}}_K}} )}^H} \cdot ( {{{\mathbf{X}}^T}{{\mathbf{G}}^T}{{\mathbf{\Phi }}^T} \otimes {{\mathbf{I}}_K}} )} \right\} \nonumber\\
&= \frac{{2T}}{{\sigma _{\text{R}}^2}}{\text{Re}}\left\{ {( {{{\mathbf{\Phi }}^*}{{\mathbf{G}}^*}{{\mathbf{R}}_{\text{X}}^*}{{\mathbf{G}}^T}{{\mathbf{\Phi }}^T}} ) \otimes {{\mathbf{I}}_K}} \right\},\\
{{{\mathbf{\tilde F}}}_{{{\mathbf{h}}_{\text{I}}}{{\mathbf{h}}_{\text{R}}}}} &=  - {{{\mathbf{\tilde F}}}_{{{\mathbf{h}}_{\text{R}}}{{\mathbf{h}}_{\text{I}}}}} \nonumber\\
&= \frac{{2T}}{{\sigma _{\text{R}}^2}}{\text{Im}}\left\{ {{{( {j{{\mathbf{X}}^T}{{\mathbf{G}}^T}{{\mathbf{\Phi }}^T} \otimes {{\mathbf{I}}_K}} )}^H} \cdot ( {j{{\mathbf{X}}^T}{{\mathbf{G}}^T}{{\mathbf{\Phi }}^T} \otimes {{\mathbf{I}}_K}} )} \right\} \nonumber\\
&= \frac{{2T}}{{\sigma _{\text{R}}^2}}{\text{Im}}\left\{ {( {{{\mathbf{\Phi }}^*}{{\mathbf{G}}^*}{{\mathbf{R}}_{\text{X}}^*}{{\mathbf{G}}^T}{{\mathbf{\Phi }}^T}} ) \otimes {{\mathbf{I}}_K}} \right\}.
\end{align}

\section*{Appendix C}
\section*{Proof of Proposition 4}
\label{AppendixC}
By introducing a new auxiliary variable $\varsigma \in \mathbb{R}$, we have
\begin{align}
	\label{new_NK}
{\tilde N} = \frac{{{Q_{{\text{tot}}}}}}{{\left( {1 + \varsigma } \right){W_{\text{I}}}}}, {\tilde K} = \frac{{\varsigma {Q_{{\text{tot}}}}}}{{\left( {1 + \varsigma } \right){W_{\text{s}}}}}.
\end{align}
We can use $f_1\left( \varsigma  \right)$ as a function w.r.t. $\varsigma$ to present ${{\tilde N}^2}( {{{\tilde K}^3} - {\tilde K}} )$. For $f_1\left( \varsigma  \right)$, it can be derived as 
\begin{align}
	f_1 \left( \varsigma  \right) =& \frac{{Q_{{\text{tot}}}^2}}{{{{\left( {1 + \varsigma } \right)}^2}W_{\text{I}}^2}}\left( {\frac{{{\varsigma ^3}Q_{{\text{tot}}}^3}}{{{{\left( {1 + \varsigma } \right)}^3}W_{\text{s}}^3}} - \frac{{\varsigma {Q_{{\text{tot}}}}}}{{\left( {1 + \varsigma } \right){W_{\text{s}}}}}} \right) \nonumber\\
	=& \frac{{Q_{{\text{tot}}}^3}}{{W_{\text{I}}^2W_{\text{s}}^3}}\frac{{\left( {Q_{{\text{tot}}}^2{\varsigma ^3} - W_{\text{s}}^2\varsigma {{\left( {1 + \varsigma } \right)}^2}} \right)}}{{{{\left( {1 + \varsigma } \right)}^5}}}.
\end{align}
The first-order derivative of $f\left( \varsigma  \right)$ w.r.t. $\varsigma$ is given by
\begin{align}
	\label{first-order_opt}
	\frac{{df\left( \varsigma  \right)}}{{d\varsigma }} = \frac{{Q_{{\text{tot}}}^3\left( {Q_{{\text{tot}}}^2 + W_{\text{s}}^2} \right)f_2 \left( \varsigma  \right)}}{{W_{\text{I}}^2W_{\text{s}}^3{{\left( {1 + \varsigma } \right)}^6}}},
\end{align}
where $f_2 ( \varsigma  ) =  { {\beta _4}{\varsigma ^3} + 3{\varsigma ^2} + {\beta _5}} $. By further taking the first-order derivative of $f_2 ( \varsigma  )$ w.r.t. $\varsigma$, we obtain $\frac{{d f_2 \left( \varsigma  \right)}}{{d\varsigma }} = 3{\beta _4}{\varsigma ^2} + 6 \varsigma$. When $0 \textless \varsigma \textless -2/{\beta _4}$, we have $\frac{{d f_2 \left( \varsigma  \right)}}{{d\varsigma }} \textgreater 0$, i.e., $f_2 ( \varsigma  )$ monotonically increases with $\varsigma$. When $\varsigma \textgreater -2/{\beta _4}$, we have $\frac{{d f_2 \left( \varsigma  \right)}}{{d\varsigma }} \textless 0$, i.e., $f_2 ( \varsigma )$ monotonically decreases with $\varsigma$. Since $f_2 ( -2/{\beta _4} ) \textgreater 0$, there exists a single unique root $\varsigma^\text{rt} \in \left( -2/{\beta _4}, +\infty \right)$ for \eqref{first-order_opt}, which is given by $\varsigma^\text{r1} \!=\! - 1/{\beta _4}  \!+\! \sqrt[3]{{( {\beta _6} \!+\! {\beta _7})/(2\beta _4^3)}} \!+\! \sqrt[3]{{( {\beta _6} \!-\! {\beta _7})/(2\beta _4^3)}}$. When $-2/{\beta _4} \textless \varsigma \textless \varsigma^\text{r1}$, we have $f_2 \left( \varsigma  \right) \textgreater 0$ and $\frac{{d f_1 \left( \varsigma  \right)}}{{d\varsigma }} \textgreater 0$, i.e., $f_1 \left( \varsigma  \right)$ monotonically increases with $\varsigma$. When $\varsigma \textgreater \varsigma^\text{r1}$, we have $f_2 \left( \varsigma  \right) \textless 0$ and $\frac{{d f_1 \left( \varsigma  \right)}}{{d\varsigma }} \textless 0$, i.e., $f_1 \left( \varsigma  \right)$ monotonically decreases with $\varsigma$. Hence, $f_1 \left( \varsigma  \right)$ is maximized at $\varsigma^\text{opt} = \varsigma^\text{r1}$. Substituting $\varsigma^\text{r1}$ into \eqref{new_NK}, the optimal solution is given by \eqref{opt_NK}. 

${{W_{\text{s}}^2}}/{{( {Q_{{\text{tot}}}^2 + W_{\text{s}}^2} )}}$ can be approximated as 0 when the total power/cost/number is sufficient. In this case, $f_2 ( \varsigma )$ is rewritten as $f_2 ( \varsigma  ) = { {\beta _4}{\varsigma ^3} + 3{\varsigma ^2}}$ with the root $\varsigma^\text{r2} = -3/{\beta _4}$. When $0 \textless \varsigma \textless \varsigma^\text{r2}$, we have $f_2 ( \varsigma ) \textgreater 0$ and $\frac{{d f_1 \left( \varsigma  \right)}}{{d\varsigma }} \textgreater 0$, i.e., $f_1 ( \varsigma  )$ monotonically increases with $\varsigma$. When $\varsigma \textgreater \varsigma^\text{r2}$, we have $f_2 \left( \varsigma  \right) \textless 0$ and $\frac{{d f_1 \left( \varsigma  \right)}}{{d\varsigma }} \textless 0$, i.e., $f_1 \left( \varsigma  \right)$ monotonically decreases with $\varsigma$. Hence, $f_1 \left( \varsigma  \right)$ is maximized at $\varsigma^\text{sub-opt} = \varsigma^\text{r2}$ and the CRB is minimized. Substituting $\varsigma^\text{r1}$ into \eqref{new_NK}, the sub-optimal number of IRS elements and sensors are given by \eqref{subopt_NK}. Thus, we complete the proof.

\bibliographystyle{IEEEtran}
\bibliography{refs.bib} 

\end{document}